# Updated View on Titan's Habitability: Interior and Surface Aqueous Environments

Diogo Gonçalves[1], Lígia F. Coelho[2], Catarina Miranda[1], Zita Martins[1,*]

[1]Centro de Química Estrutural, Institute of Molecular Sciences, and Department of Chemical Engineering, Instituto Superior Técnico, Universidade de Lisboa, Av. Rovisco Pais 1, 1049-001 Lisbon, Portugal

[2]Department of Astronomy and Cornell Center for Astrophysics and Planetary Science, Cornell University, Ithaca, NY, 14850, USA.

*Corresponding author (zita.martins@tecnico.ulisboa.pt)

## Abstract

Saturn's moon Titan is the only world beyond Earth where abundant organic molecules and liquid water are expected to interact. Crucially, these interactions yield prebiotic molecules central to terrestrial biochemistry. Characterizing this prebiotic chemistry is one of the main science goals of NASA's upcoming Dragonfly mission. Here we synthesize the current state of knowledge on the habitability and prebiotic potential of Titan's three aqueous settings: its interior hydrosphere, its putative cryovolcanic flows, and its impact-melt pools. For each setting, we identify the enabling factors and the principal bottlenecks that limit prebiotic complexification. Titan's interior is likely pervaded by liquid water in contact with organic material delivered from the surface and released from the core, though the longevity of its liquid water has recently been called into question. Cryovolcanic regions transiently expose surface organics to ammonia-bearing aqueous melts but are constrained by short freezing timescales. Impact-melt pools, by contrast, are warm and long-lived environments whose prebiotic chemistry is limited by the scarcity of ammonia. Dragonfly will return ground-truth data not only for Titan's habitability, but also for organic-rich aqueous environments across the Solar System and beyond. The years before its arrival are the last in which predictive work can still shape Dragonfly's investigations.

## 1. Introduction

Saturn's largest moon Titan is the only natural satellite in the Solar System with a substantial atmosphere, at ~1.5 bar surface pressure, composed mostly of molecular nitrogen ($N_2$; ~95%) and methane ($CH_4$; ~5%) (Hörst 2017; Nixon 2024). Solar ultraviolet (UV) photons and energetic particles from Saturn's magnetosphere dissociate atmospheric methane and nitrogen to drive intermingled photochemical processes. This photochemistry yields small hydrocarbon, nitrile, and oxide molecules – e.g., ethane ($C_2H_6$), acetylene ($C_2H_2$), hydrogen cyanide (HCN), and carbon monoxide (CO) (Hörst 2017; Nixon 2024) – which also serve as intermediates in the synthesis of the higher-molecular-weight species that ultimately constitute the complex organic aerosols of Titan's haze (Lebonnois et al. 2002; Wilson and Atreya 2003). All these organic products tend to sediment and accumulate as an organic layer on Titan's water-ice crust (Janssen et al. 2016; Lopes et al. 2020).

With a surface temperature of 94 K (Cottini et al. 2012a; Jennings et al. 2019), Titan also sustains a methane cycle analogous to Earth's water cycle, dynamically maintaining stable lakes and seas (Hayes et al. 2018). These liquid bodies occupy only a small fraction of a geomorphologically diverse surface, which also includes vast organic dune fields, mountainous ice terrains, fluvial networks, impact craters, and putative cryovolcanic constructs (Lopes et al. 2020; Schoenfeld et al. 2023). Beneath Titan's crust lies a differentiated interior comprising a hydrated rocky core, high-pressure ice layers, and a global liquid-water ocean (Sotin et al. 2021) – though the decades-old consensus on the latter is currently being challenged (Petricca et al. 2025). Moreover, Titan's core is increasingly understood to have accreted from organic-rich material (Miller et al. 2019; Néri et al. 2020), making Titan organic-rich at depth as well as at the surface.

Indeed, Titan's organic richness is one of its most striking features. Compounding this organic richness is a second feature that underpins much of Titan's astrobiological interest: liquid water is expected to interact with this plethora of organic material across interior and surface settings. In the interior, organics may be transferred to the liquid hydrosphere from the organic-rich surface above and the rocky core below, raising the prospect of a subsurface pervaded by water–organic interactions (Lunine et al. 2020). At the surface, Titan's organic layer can also be brought into aqueous phase whenever liquid water exists transiently on the surface, either in putative ammonia–water cryovolcanic flows erupted from Titan's interior (Mitri et al. 2008; Lopes et al. 2013) or in impact-generated melt pools (Solomonidou et al. 2020; Hedgepeth et al. 2022). Crucially, laboratory simulations of these water–organic interactions have consistently produced prebiotic molecules including amino acids, nucleobases, and related feedstocks (Neish et al. 2018; Farnsworth et al. 2024), fundamental for life as we know it on Earth.

A second source of astrobiological interest in Titan has been the possibility of it harboring hydrocarbon-based life (McKay 2016; Lv et al. 2017; Lunine et al. 2020). Although the thermal energy available at ~94 K severely limits life-like processes and molecular interactions (Lunine et al. 2020; Bains et al. 2024), a growing body of work shows that Titan's cryogenic hydrocarbon environments support, at least, a rich supramolecular chemistry. Atmospherically deposited organics can form a diverse array of co-crystals on Titan's surface (e.g., Thakur and Remsing 2023, 2024; Czaplinski et al.

2025; Francis et al. 2026), which may trace past cryovolcanic, fluvial, or degassing activity, embody organic diagenesis, or concentrate prebiotic ingredients and metabolic energy sources (Cable et al. 2021; Czaplinski et al. 2025). Recently, Izquierdo-Ruiz et al. (2025) reported stable co-crystals of nonpolar methane and ethane with highly polar hydrogen cyanide, suggesting that prebiotically relevant polar molecules may be more available in Titan's hydrocarbon lakes than previously assumed. This conclusion was supported by Vu and Hodyss (2026), who showed that acrylonitrile – the proposed building block of "azotosome" membranes (Stevenson et al. 2015) – forms a stable co-crystal with ethane up to 166 K. Their results, however, weaken the thermodynamic case for spontaneous vesicle self-assembly on Titan's hydrocarbon lakes (Sandström and Rahm 2020), though Mayer and Nixon (2025) argue that this thermodynamic barrier could be bypassed through an aerosol-droplet formation mechanism. These research avenues would be informed by a range of instrumentation and mission concepts that continue to be proposed, aiming to characterize Titan's hydrocarbon lakes and streams (e.g., Stofan et al. 2013; Lorenz et al. 2018a; Tagliabue et al. 2020; Mitri et al. 2021; Martins et al. 2024; Gonçalves et al. 2026). Such characterization would also inform astrobiology-related questions tied to fluvial erosion and transport, or to the percolation of hydrocarbons into the ice crust, as detailed below. Hydrocarbon-based life, however, is considered unlikely under Titan's surface conditions (Lunine et al. 2020), and we do not pursue it further here.

We instead focus on the sites of potential aqueous habitability on Titan (**Figure 1**), in line with the main science goals of the National Aeronautics and Space Administration's (NASA) upcoming Dragonfly mission (Barnes et al. 2021). Scheduled to launch in 2028 and arrive at Titan in 2034, Dragonfly will substantially build on the Cassini–Huygens mission (Lopes et al. 2025), a joint NASA, European Space Agency (ESA), and Italian Space Agency (ASI) endeavor that explored the Saturnian system from 2004 to 2017. By delivering a relocatable rotorcraft lander to Titan's surface (Lorenz et al. 2018b), Dragonfly will take advantage of the moon's dense atmosphere, relatively weak winds, and low gravity to sample its organic sediments, water-ice bedrock, and impact-melt deposits. Throughout its traverse, the Dragonfly mission will assess Titan's prebiotic chemistry, habitability, and potential biosignatures (Lorenz et al. 2021; Barnes et al. 2021). It will employ a science payload that includes a mass spectrometer (DraMS; Grubisic et al. 2021; Moulay et al. 2023; Boulesteix et al. 2024; Freissinet et al. 2024), a gamma-ray and neutron spectrometer (DraGNS; Burks et al. 2020; Lawrence et al. 2022), a geophysics and meteorology package (DraGMet; Chatain et al. 2023; Mortensen et al. 2024a, b; Wray et al. 2024) and a camera suite (DragonCam; Clark et al. 2025).

After a multi-site traverse extending over 100 km, the Dragonfly prime campaign will culminate inside Selk crater, where a post-impact water melt – now frozen – is expected to have enabled liquid water–organic interactions in the past (Neish et al. 2018; Lorenz et al. 2021; Barnes et al. 2021). Although Selk will be the only water–organic environment directly sampled by Dragonfly, it will constrain the prebiotic chemistry expected wherever else on Titan organics and liquid water interact. On Titan, the prebiotic chemistry of water–organic interactions can serve as a proxy for habitability. A habitable environment is one that supplies bioessential elements (CHNOPS), a solvent (notably liquid water), and accessible energy sources, under conditions tolerable to known biochemistry (Cockell et al. 2016, 2024). Across Titan's aqueous environments, the co-occurrence of those factors should correlate with the production of prebiotic molecules. With this correlation in mind, we review the current state of knowledge on the habitability and prebiotic potential of the environments on Titan that could harbor liquid water. Section 2 reviews the long-standing consensus on Titan's internal global water ocean and the habitability framework built upon it, before turning to recent modeling work on material delivery to Titan's interior hydrosphere. We also review the "mushy" interior model proposed by Petricca et al. (2025) and discuss how it may affect the perceived habitability of Titan's interior, in the hope of motivating dedicated studies building on this new model. Section 3 turns to the surface and reviews its material and energy influxes, informing the habitability of Titan's cryovolcanic regions and impact craters – two settings where liquid water may transiently exist. Sections 4 and 5 develop the specifics of each surface environment, treating their prebiotic potential and how Dragonfly may sample it. Section 6 discusses how the findings reviewed here inform the habitability and prebiotic potential of analogous environments across the Solar System and beyond.

## 2. Titan's interior

Over its 13-year orbital tour, the Cassini spacecraft performed nine close Titan flybys dedicated to gravity-field determinations (Iess et al. 2010, 2012; Durante et al. 2019). One of the most consequential inferences of this campaign was the presence of a deep, global liquid water ocean decoupled from the solid interior (Iess et al. 2012; Mitri et al. 2014b; Sotin et al. 2021). Gravity measurements further constrained Titan's moment of inertia, indicating that it is internally differentiated into an approximately 600 km-thick hydrosphere overlying a low-density rocky core that holds hydrated silicates (Iess et al. 2010; Castillo-Rogez and Lunine 2010) and a substantial fraction of organic material (Néri et al. 2020).

Titan was, thus, uncontroversially classified as an ocean world (Lunine 2017; Hendrix et al. 2019). Under that prevalent model, the study of the habitability of Titan's interior revolved around whether chemical elements and energy sources could reach its ocean at sufficient rates to nurture a biosphere. Two delivery pathways have been proposed, pathways which remain pertinent today and acquire new significance under the recent reinterpretation of Titan's hydrosphere (section 2.1). The first is sourced from the surface, whose plethora of organic molecules are delivered to the ocean by the action of impacts (Crósta et al. 2021; Neish et al. 2024; Kalousová et al. 2024, section 2.2). The second originates from the rocky core: accreted from carbonaceous material, it has been progressively releasing organic material into the hydrosphere (Miller et al. 2019, 2026; Reynard and Sotin 2023, section 2.3). Together, these two pathways have raised the possibility that geochemical input of complex organics could enable a heterotrophic biosphere in Titan's ocean (Affholder et al. 2025).

### 2.1. The interior structure: from a global ocean to a mushy hydrosphere

Titan's ocean-world classification rested on a structure comprising a global subsurface ocean sandwiched between an outer ice shell and a high-pressure ice layer (**Figure 1a**). The outermost layer is composed by water ice $I_h$ and, likely, methane clathrate hydrates at shallow depths (Tobie et al.

2005; Lunine and Lorenz 2009). The thickness of the outer ice shell is only loosely constrained, with estimates spanning 50–200 km (Hemingway et al. 2013; Mitri et al. 2014b; Vance et al. 2018), and the extent of convection within it still unclear (Kalousová and Sotin 2020). At the base of Titan's hydrosphere, water ice stabilizes in high-pressure phases. The thickness of this high-pressure ice layer was estimated at 50–300 km, with the uncertainty driven largely by the poorly constrained density of both the core and the ocean (Journaux et al. 2020). Generally considered a near-total barrier to chemical exchange between the ocean and the rocky core (Hand et al. 2020), it was, nevertheless, recognized that convection within it could transport material upward to the ocean (Choblet et al. 2017; Journaux et al. 2017; Kalousová and Sotin 2018): if heat flux or salinity were sufficiently high, the base of the high-pressure layer would remain partially molten, allowing material transfer from the core to the overlying ocean (Journaux et al. 2020).

Titan's ocean, decoupling both ice layers, was considered substantially denser than pure water by roughly 5–10% (Sotin et al. 2021). Based on that, its proposed composition included solutes like ammonium sulfate (Fortes et al. 2007), sodium chloride, and magnesium sulfate (Vance et al. 2018; Leitner and Lunine 2019). This understanding evolved with a reanalysis by Goossens et al. (2024), which, using an improved treatment of tidal effects, estimated a substantially lower ocean density (~1,091 kg $m^{-3}$) than earlier estimates (~1,300 kg $m^{-3}$; e.g., Mitri et al. 2014b; Baland et al. 2014). At this revised density, a water or ammonia–water ocean would be more likely (Goossens et al. 2024), disregarding the need for large quantities of dense solutes. Overall, the geochemical state of the ocean has remained uncertain (Sotin et al. 2021; Nixon et al. 2025). Nevertheless, it was believed that such geochemical state enabled the ocean to remain liquid over Titan's ~4.5 Gyr history (Lunine et al. 2020 and references therein). The liquid state was sustained by the heat generated from the decay of radiogenic elements – mostly $^{40}$K and $^{235}$U – and tidal friction due to the gravitational interaction with Saturn, freezing-point depression by its solutes, and the insulating effect of the outer ice shell (Sotin et al. 2021).

The existence of a liquid subsurface layer was supported by four independent lines of evidence: an anomalously large obliquity, the degree-2 tidal Love number $k_2$, early measurements of a non-synchronous rotation, and an apparent Schumann-like electromagnetic resonance detected by the Huygens probe. Of these, only the obliquity and $k_2$ evidence survived subsequent scrutiny (Lorenz and Le Gall 2020; Sotin et al. 2021). At the same time, Titan's obliquity – three times the value previously predicted for a fully solid body (Bills and Nimmo 2008), and therefore taken as evidence for an internal fluid layer (Baland et al. 2011, 2014) – was also reproduced by large solid-body tidal deformations (Baland et al. 2016). The measured obliquity was thus deemed a necessary but not sufficient condition for an interior ocean. The tidal Love number $k_2$ remained the primary observational evidence of the ocean hypothesis. This evidence was, however, recently challenged by Petricca et al. (2025), following measurements of an unexpectedly large imaginary component of $k_2$ (Downey and Nimmo 2025). These measurements entailed tidal energy dissipation rates 3–4 times larger than the maximum expected for a body with a subsurface ocean. Since a liquid layer would suppress tidal dissipation in the layers beneath it, no model incorporating a global ocean could reproduce these novel measurements (Petricca et al. 2025). According to the authors, the strong tidal dissipation in the interior of Titan – also inferred by Downey and Nimmo (2025) – is concentrated within the high-pressure ice layer above the core. To sustain dissipation at the inferred rate, the layer must deform with a viscosity only attainable when close to its melting point, leading to the proposal that Titan's interior consists of a mushy hydrosphere, made of high-pressure ice interspersed with melt pockets (**Figure 1a**). Aided by fractional crystallization, the proposed melt pockets can constitute nutrient-rich sites with concentrated salts, volatiles, and dissolved organics. Populating most of the hydrosphere, they can in total contain as much liquid water as Earth's oceans (Petricca et al. 2025). This reinterpretation of Titan's interior also underlines that the high-pressure ice layer is not a chemical barrier but a conduit of chemical elements and energy sources (Kalousová and Sotin 2018; Journaux et al. 2020), which, originating from the core, are likely dispersed by vigorous convection throughout Titan's hydrosphere (Kalousová and Sotin 2020).

The predicted convective transport does not denote, however, hydrothermal circulation within the rocky core itself. In any planetary body where a hydrosphere overlies a rocky core, the flux of redox-active species delivered into the hydrosphere depends on the degree to which liquid water interacts with the rocky core. This interaction remains fundamentally limited on Titan, given its high hydrostatic pressure at the ice-rock interface. Estimated at ~1.1 GPa, the hydrostatic pressure shall close the rock pore spaces within a few hundred meters of the rock surface (Byrne et al. 2018) – a hypothesis consistent with total pore closures of chondritic materials being measured at 360 MPa (Seltzer et al. 2024) –, effectively precluding any sustained water percolation. This contrasts with Enceladus and Europa, where direct seafloor water-rock interaction is expected to generate the chemical disequilibria necessary to sustain chemoautotrophic life. Titan's hydrosphere is unlikely to harbor abundant inorganic redox couples of hydrothermal origin. Any biosphere persisting within its interior would more plausibly depend on organic-based energy sources. The question then becomes whether the organic inventory that reaches Titan's interior is sufficient to sustain such biosphere.

### 2.2. Organic delivery from the surface

Titan's surface is unparalleled in the solar system for its sustained inflow of atmospheric organic material. If part of this material reaches the interior liquid water, it could hypothetically sustain a heterotrophic biosphere (Yanez et al. 2024; Affholder et al. 2025). However, the thick outer ice shell significantly hinders this surface-to-interior organic delivery (Neish et al. 2024). *Three* delivery mechanisms have been proposed to overcome it. The *first* is the direct breaching of the ice shell by a large impactor. Crósta et al. (2021) modelled the formation of Menrva, Titan's largest known impact crater at ~400 km diameter and found that it likely penetrated the outer ice shell, transiently mixing organic-rich surface material with interior liquid water. Crósta et al. (2021) further suggested that similar breaching may have occurred for craters in the 70–120 km range, though this has not been directly modelled. On the other hand, modeling by Wakita et al. (2023) suggested that even a 4 km impactor produces transient excavation cavities only 15–20 km deep, insufficient to reach the ocean through a thick ice shell. Indeed, Menrva stands out as exceptionally large within Titan's crater population (Hedgepeth et al. 2020; Rossignoli et al. 2022), underscoring, by extension, the rarity of its putative breaching event. The *second* delivery mechanism is the sinking of impact melt pools through the ice, driven by their higher density relative to the surrounding ice (Neish et al. 2024; Kalousová et al. 2024). Kalousová et al. (2024) showed that this process succeeds only under a combination of conditions: a methane clathrate layer at least 10–15 km thick – insulating the ice below and keeping it warmer and less viscous –, combined with either low ice viscosity, high surface porosity, or internal tidal heating. The *third* suggested delivery mechanism consists of transport by solid-state convection within the outer ice shell (Mitri and Showman 2008; Kalousová and Sotin 2020; Sotin et al. 2021). In this scenario, surface organics would need to be tectonically buried to depths beyond the brittle–ductile transition, estimated at ~12–45 km (Schurmeier et al. 2023), where the ice shell may thermally convect on overturn timescales of $10^4$–$10^5$ years (Mitri and Showman 2008). This pathway has been the least constrained of the three. Still, Dragonfly's seismological measurements are expected to provide direct constraints on Titan's tectonic activity (Barnes et al. 2021), constraining in turn the convective delivery of surface material to the interior.

Glycine fermentation has been adopted to evaluate the extent of the heterotrophic biosphere such surface delivery could nurture. The maximum expected flux of glycine to Titan's ocean would change its concentration in the bulk hydrosphere at a rate of only ~10 nmolal $Gyr^{-1}$ (Neish et al. 2024; Affholder et al. 2025). In what may already be an overestimation (Kalousová et al. 2024), this rate would be largely insufficient for biomass productivity, and thus to nurture a detectable global biosphere (Affholder et al. 2025). Still, the melt pockets predicted to exist within the high-pressure ice hydrosphere (Petricca et al. 2025) could harbor cell densities far exceeding the bulk-dilution modelled by Affholder et al. (2025). Adding to that, glycine is unlikely to be the only fermentable molecule delivered from the surface. Acetylenotrophy – the anaerobic degradation of acetylene – is the best-developed alternative candidate (Yanez et al. 2024), with its end products (ethanol, acetate, and hydrogen) putatively supporting secondary metabolic communities more complex and productive than glycine fermentation alone (Yanez et al. 2024). To better constrain the possibility of heterotrophic biospheres on Titan's interior, estimates of the biomasses sustained by other fermentation metabolisms – and accounting for much smaller liquid water volumes – would be valuable (Affholder et al. 2025).

It must be kept in mind, though, that the thick ice shell – most recently estimated at up to ~200 km (Goossens et al. 2024; Petricca et al. 2025) – will always severely limit the flux of surface organics to the interior of Titan. Any delivery that does occur should be episodic and spatially localized (Neish et al. 2024), raising significant hurdles for the maintenance of any sustained habitability. A potentially more impactful reservoir of organic material resides in Titan's rocky core, as discussed in the following section.

### 2.3. Material delivery from the rocky core

Carbonaceous chondrites are broadly accepted as good compositional proxies for Titan's rocky core (Fortes 2012; Tobie et al. 2012; Glein 2015; Néri et al. 2020), with multiple lines of evidence converging on a substantial insoluble organic matter (IOM) mass fraction within it. Miller et al. (2019) used chondritic and cometary data to constrain the composition of refractory organics in Titan's core and showed that thermal decomposition of these organics could have contributed approximately 50% of Titan's present-day atmospheric molecular nitrogen. Néri et al. (2020) found that a Titan interior model based on a CI carbonaceous chondrite composition alone is too dense to match the observed mass and reduced moment of inertia, and that reconciling them requires adding ~17–23 ± 10 wt% of refractory carbonaceous matter, derived from IOM, to the rocky core. Reynard and Sotin (2023) incorporated the thermodynamic effects of organic material on interior heating and independently predicted ~22 wt% complex organic material in Titan's present-day core. Dunaeva et al. (2025) constructed interior models based on both carbonaceous (CI and CM) and ordinary (L and LL) chondrite compositions and found that organic matter fractions of 10–22 wt% and 20–28 wt%, respectively, are required to satisfy Titan's mass and moment of inertia constraints. The resulting picture is of a moon accreted from organic-rich material, currently holding an interior organic reservoir of ~$10^{19}$ kg (Miller et al. 2026), orders of magnitude larger than the organic input to Titan's interior from surface impact melt delivery over 4.5 Gyr, estimated at ~$10^{16}$ kg (Neish et al. 2024; Miller et al. 2026).

The accreted organic material is thought to reach Titan's hydrosphere after thermal energy – resulting from radioactive decay and tidal dissipation (Tobie et al. 2005; Castillo-Rogez and Lunine 2010; Reynard and Sotin 2023) – decomposes the accreted IOM. As temperatures in the core rise, refractory IOM undergoes pyrolysis to produce mobile organic compounds, including volatiles that ultimately contribute to the atmosphere (Miller et al. 2019) and less volatile molecules that may partition into convective melt pockets (Kalousová and Sotin 2018; Journaux et al. 2020; Petricca et al. 2025). Also encapsulating inorganic compounds, these melt pockets could disseminate the accreted material throughout Titan's hydrosphere. Whether the degraded IOM products would be sufficient to nurture fermentative metabolisms remains an open question (Miller et al. 2026).

Finally, the same chondritic-and-cometary material that supplied this organic reservoir also very likely delivered all six CHNOPS elements: C, H, N, and O are intrinsic to the refractory organic macromolecules and to the co-accreted ices (Hersant et al. 2008), while P and S likely accreted as phosphine ($PH_3$) and hydrogen sulfide ($H_2S$) clathrates, respectively (Hersant et al. 2008; Pasek et al. 2011). The six bioessential elements are therefore expected to be primordially sequestered in Titan's core and available for release into the overlying hydrosphere on geological timescales.

Overall, the habitability of Titan's interior appears to rest primarily on its accreted material being progressively mobilized into the hydrosphere. Even so, a fundamental question remains concerning the longevity of its liquid water. The intense tidal dissipation currently observed is tied to Titan's orbital eccentricity, which should damp to zero on a timescale of 30–65 Myr (Downey and Nimmo 2025; Petricca et al. 2025), lending the same longevity to the mushy interior state. This is a striking departure from the multi-Gyr lifetime previously attributed to a global subsurface ocean (Lunine et al. 2020). Thus, although a high-pressure ice interior interspersed with melt niches may increase the habitability prospects of Titan's interior by concentrating the solutes to chemically meaningful levels, it could do so only over a geologically brief interval.

### 2.4. Probing the interior from the surface

Independently of the liquid water lifetime, any extant biosphere in Titan's interior remains buried beneath hundreds of kilometers of ice, inaccessible to direct sampling by any foreseeable surface mission. Effusive cryovolcanism nevertheless remains a possible avenue for sampling the interior's chemical composition, as will be discussed in section 4.1. Should it be demonstrated, the characterization of erupted ocean material would lie beyond the sensitivity of IR reflectance spectroscopy and would instead demand *in situ* mass spectrometry or fluorescence (Johnson et al. 2025); fortunately, Dragonfly will carry both within DraMS and DragonCam (UV fluorescence; Clark et al. 2025), respectively. Regardless, Titan's atmospheric composition also provides an indirect window into its interior. Outgassed species carry isotopic and elemental signatures that encode the history of ocean-atmosphere exchange and the thermal evolution of the rocky core (Tobie et al. 2006; Glein 2015; Miller et al. 2019). The Huygens GCMS instrument detected both primordial $^{36}$Ar and radiogenic $^{40}$Ar in Titan's atmosphere during its 2005 descent (Niemann et al. 2005; Das et al. 2025), providing direct evidence that potassium-bearing material in the interior has degassed into the atmosphere. However, heavier noble gases, including krypton and xenon, were not detected despite being actively sought (Niemann et al. 2005). Going forward, measurements of the xenon isotope abundances would place particularly strong constraints on the organic composition of the interior (Miller et al. 2026) and better inform the habitability of Titan's hydrosphere. DraMS is well positioned to make these measurements (Barnes et al. 2021). Complementing these atmospheric measurements, DraGMet will constrain the density and rheological structure of Titan's hydrosphere, testing whether a global ocean is present or whether an ocean-free model better describes Titan's interior. DraGMet may also estimate the thickness of the outer ice shell and the degree of tectonic activity within it (Delaroque et al. 2026), thus informing the organic flux that the outer shell permits.

Terrestrial environments that share key physical and chemical characteristics with Titan's hydrosphere can serve as laboratories for developing and testing further habitability hypotheses (Nixon et al. 2026). Fermentation-based communities in energy-limited settings are particularly relevant given the metabolic framework proposed for Titan's melt niches. A relevant analogue is the deep crystalline bedrock fracture water system of the Fennoscandian Shield, where heterotrophic communities dominated by fermentative microorganisms are sustained by ancient organic carbon, under extremely low energy fluxes, and with biomass turnovers on geological timescales (Purkamo et al. 2014). Such systems demonstrate that fermentative metabolisms can persist over geological timescales in isolated, low-flux environments (Bradley et al. 2020), like Titan's hydrosphere with its primordial organic inventory. Applying quantitative bioenergetic frameworks (e.g., Affholder et al. 2025) to these systems represents a productive path for constraining biomass estimates on Titan's hydrosphere and informing the selection of biosignatures and life detection strategies for Dragonfly.

## 3. Titan's surface

Titan's surface is geologically young by Solar System standards. Early estimates of Titan's crater retention age (i.e., the duration over which craters remain detectable in Cassini data) placed it at 200–1,000 Myr (Neish and Lorenz 2012; Hedgepeth et al. 2020). These earlier estimates relied on crater scaling laws derived from pure water-ice or sand targets, which may not adequately represent the composition of Titan's upper crust. Instead, Titan's upper crust is likely composed of methane clathrate hydrates (Sotin et al. 2021) – crystalline water-ice structures that encage methane or ethane molecules. They are thermodynamically stable under Titan's surface conditions (Tobie et al. 2006; Choukroun et al. 2010; Vu et al. 2020) and should readily form when hydrocarbons percolate into the porous crust (Choukroun et al. 2010; Vu et al. 2020) or ascend from the interior (Tobie et al. 2006; Carnahan et al. 2022). Crucially, because methane clathrates have a significantly lower thermal conductivity than water ice, they produce steeper near-surface temperature gradients (Kalousová & Sotin 2020), which in turn affect the crust's yield strength and, ultimately, the impact crater scaling (Wakita et al. 2026). Assuming that Titan's crust is capped by a methane clathrate layer, Wakita et al. (2026) estimated a crater retention age of 300–340 Myr, ruling out ages older than 500 Myr. Ultimately, this revised estimate further narrows the chemical history accessible at Titan's surface.

In contrast to its interior, Titan's surface is an open system whose material (section 3.1) and energy (section 3.2) inputs have been directly characterized by the Cassini–Huygens mission (Tomasko et al. 2005; Niemann et al. 2010; Das et al. 2025). These inputs feed the transient aqueous environments – cryovolcanic flows (section 4) and impact melt-pools (section 5) – whose habitability and prebiotic potential we discuss later. Although we do not expect to find liquid-water environments on the current surface of Titan, the now-frozen cryovolcanic and impact regions should still preserve the products of their habitable, chemically active past.

### 3.1. Atmospheric delivery of organics and CHNOPS

Over geological time, the sedimentation of the photochemical products – particularly the macromolecular haze – has built up a global organic-rich sediment layer (Janssen et al. 2016) that continues to accumulate at a rate of approximately 3.5 kg $cm^{-2}$ $Gyr^{-1}$ (Neish et al. 2024). Since the late 1970s, there has been a sustained effort to replicate Titan's atmospheric haze particles through laboratory experiments. In these, vacuum vessels filled with varying mixtures of nitrogen, methane, and other gases are subjected to energetic stimulation by UV radiation or electrical discharge (Cable et al. 2012; Coll et al. 2013; Hörst 2017; Nixon 2024), yielding organic precipitates known as tholins (Sagan and Khare 1979). Ongoing studies still seek to better understand the precipitates' formation mechanisms (e.g., Dubois et al. 2020; Carrasco et al. 2022; Ugelow et al. 2024; Couturier-Tamburelli et al. 2025) and composition (e.g., He et al. 2022; Sciamma-O'Brien et al. 2023; Yang et al. 2025; Jansen et al. 2025) as well as to standardize their characterization among laboratories (Li et al. 2022).

Extending these experiments, Hörst et al. (2012) demonstrated that the inclusion of CO in the initial gas mixtures produces biologically relevant molecules, including amino acids and nucleobases. Sebree et al. (2018) subsequently developed a systematic analytical framework for detecting and quantifying such prebiotic species in laboratory aerosol analogs, with Pearce et al. (2024a) using the same framework to show that, at 0.2% CO concentration, a rich prebiotic inventory is produced – including at least seven nucleobases, nine proteinogenic amino acids, and five additional biomolecules such as urea. Together, these results have established that prebiotic molecules can form directly during atmospheric haze production and be delivered to the surface. Such prebiotic molecules should deposit as solid sediments, much like the nitrile species produced photochemically in the atmosphere (Yu et al. 2024). Their solid-state condensation is expected to influence their photostability during settling (Gonçalves et al. 2025b), their reactivity upon contact with surface liquids (Yu et al. 2020; Li et al. 2022), and their spectral signatures on dry surfaces (Gonçalves et al. 2025a).

However, the lowest CO concentration shown to yield prebiotic molecules (0.2%; Pearce et al. 2024a), is still more than an order of magnitude higher than the expected CO concentration in Titan's atmosphere (~47 ppm; Cottini et al. 2012b, and references therein). Consistent with that, no O-bearing organic molecules – let alone prebiotic species – have yet been detected in Titan's atmosphere (Nixon 2024). The latter's organic inventory consists mostly of C, H, and N with minimal O (de Kok et al. 2007; Hörst 2017). Still, P has now been confirmed in plume material from Enceladus (Postberg et al. 2023; Khawaja et al. 2025), with S tentatively detected as $H_2S$ (Waite et al. 2009), making it feasible that trace amounts of P (and possibly S) arrive at the top of Titan's atmosphere – as apparently occurs for O (Hörst et al. 2008) – and accumulate at the surface.

### 3.2. Energy sources at the surface

Beyond chemical feedstocks, several energy sources have been identified in Titan's atmosphere, including solar UV photons, energetic photoelectrons, magnetospheric electrons and ions, metallic ions from micrometeorite ablation, and galactic cosmic rays (Hörst 2017; Nixon 2024). These energy inputs are deposited at different altitudes, depending on the penetration depth of the relevant particles and photons. As a result, only ~10% of the solar flux reaching Titan penetrates to the surface (Read et al. 2016), amounting to a mere ~0.1% of the solar energy flux received at Earth's surface (McKay et al. 1991). Low as it may seem, it still exceeds the photon flux required to sustain photosynthesis by several terrestrial organisms (McKay 2014).

In parallel, the atmospheric photoproducts constitute a significant reservoir of chemical energy at Titan's surface. For instance, the hydrogenation reactions of acetylene and hydrogen cyanide are strongly exergonic (McKay 2016), with these molecules being among the most abundant products of the atmospheric photochemistry (Nixon 2024). The hydrogenation of acetylene to methane has a Gibbs free energy of $-334$ kJ $mol^{-1}$ (McKay and Smith 2005), a chemical disequilibrium identical to that exploited by methanogenic organisms on Earth. Intriguingly, the Huygens probe detected a drop in the molecular hydrogen mole fraction of $\sim 3 \times 10^{-5}$ over the last 3 km above the surface (Niemann et al. 2010), which, although small, could indicate consumption by some chemical or metabolic process (Strobel 2010). However, subsequent reanalyses suggest that the apparent flux may instead be an artifact arising from an error in instrumental calibration (Strobel 2022). While the recent reanalysis of the Huygens GCMS dataset (Gautier et al. 2024; Das et al. 2025) did not revisit this topic, Dragonfly will construct a vertical profile of hydrogen abundance during its flights to determine whether there is a net flux into the surface (Barnes et al. 2021) and how much chemical energy it can provide therein.

A significant limitation on the productive use of this chemical energy is the 94 K surface temperature. Because temperature controls the reaction rates exponentially, a chemical process that proceeds on microsecond timescales at ambient terrestrial temperature would require years at Titan's surface (Lunine et al. 2020). Molecular solubility is also a concern: even simple apolar molecules dissolve in liquid ethane at 94 K to concentrations three orders of magnitude below that of a modest globular protein in water at room temperature (Malaska and Hodyss 2014; Lunine et al. 2020). Transient warm environments overcome both limitations. Impact craters and putative cryovolcanic flows provide thermal energy sufficient to overcome kinetic and solubility barriers and to access the chemical energy stored in the surface organics. The energetic-budget analyses described above for a water-based heterotrophic biosphere (section 2) therefore extend naturally to these environments – and with greater force, given their direct access to the vast reservoir of fermentable organic material at the surface.

## 4. Cryovolcanism

Cryovolcanism is the eruption of liquid or vapor phases of water or other volatiles, with or without entrained solids, that would be frozen at the normal surface temperature of an icy satellite (Geissler 2015; Fagents et al. 2022). It has long been invoked as a mechanism capable of delivering subsurface liquid water to Titan's surface (Kargel 1991, 1992), though whether it remains active today is an open question (Nixon et al. 2025). The Cassini–Huygens mission could not settle this question, given the limited spatial resolution of its instruments. The Cassini Titan Radar Mapper (RADAR) instrument imaged approximately two-thirds of Titan's surface at spatial resolutions no better than 350 m and typically ~1 km per pixel (Elachi et al. 2005; Moore and Pappalardo 2011), while the Composite Infrared Spectrometer (CIRS) could not search for temperature anomalies at spatial scales below 100 km (Nixon et al. 2019), rendering it largely unsuitable for detecting cryovolcanogenic hotspots. Early interpretations nevertheless proposed a substantial inventory of candidate cryovolcanic constructs (Elachi et al. 2005; Lopes et al. 2007). Subsequent analyses, however, showed that some of these features lack the expected dome-like profile, while others – initially identified as flow-like units – turned out to be more consistent with a fluvial than an effusive origin. As a result, the inventory of likely cryovolcanic features on Titan is considerably smaller today than once thought, having remained unsettled (Lopes et al. 2013). Throughout these reassessments the Sotra Patera complex (**Figure 1b**) has remained the strongest candidate for cryovolcanic activity on Titan. This region comprises a ~17 km wide, >1 km deep non-circular depression interpreted as a collapse caldera (Sotra Patera), an adjacent mountain ~1.5 km high (Doom Mons), and extensive lobate deposits (Mohini Fluctus) that appear to emerge from Doom Mons (Lopes et al. 2013). Sohl et al. (2014) showed that Sotra Patera falls in the center of a region of heightened tidal flexure and stress concentration, lending independent geophysical support to the cryovolcanic interpretation of this region. Radiative transfer analyses based on Cassini's Visible and Infrared Mapping Spectrometer (VIMS) revealed localized brightening of the Sotra Patera region, contrasting with surrounding terrains, possibly indicative of active degassing (Solomonidou et al. 2016). Still, no geophysical model has yet fully explained how the Sotra Patera complex may have arisen (Nixon et al. 2025). More recently, a subset of depressions in the north polar region have also been interpreted as explosion craters formed by violent cryovolcanic episodes (Mitri et al. 2019; Wood and Radebaugh 2020; Brouwer et al. 2024). At the same time, uplifted and radially dissected "labyrinth" terrains have also been interpreted as the surface expressions of cryomagmatic intrusions within the ice shell (Schurmeier et al. 2023). We next review candidate formation mechanisms for these putative cryovolcanic features, detailing how each could enable contact between liquid water and Titan's organic inventory (section 4.1), their prebiotic potential (section 4.2) and their characterization by Dragonfly (section 4.3).

### 4.1. Modes of cryovolcanism and the liquid water–organic interaction they enable

Three manifestations of cryovolcanism have been discussed for Titan. The most developed of these is effusive cryovolcanism, backed by the ammonia–water cryomagma scenario of Mitri et al. (2008). It invokes cracking at the base of the ice shell and the formation of ammonia–water pockets within the ice, which, as they undergo partial freezing, progressively concentrate the dissolved ammonia toward the peritectic of the ammonia–water system (~33 wt% $NH_3$ at ~176 K; Kargel 1992). At this point, the liquid approaches neutral buoyancy with respect to water ice (Mitri et al. 2008), sufficiently so that large-scale tectonic stresses can drive effusive eruptions onto the surface, like Sotra Patera's lobate features. An intense tidal dissipation (Downey and Nimmo 2025) would sustain vigorous solid-state convection within the outer ice shell, contributing to the proposed upward transport of cryomagma. An intriguing question is whether the ammonia–water pockets envisaged by Mitri et al. (2008) could be the same as, or be fed by, the tidally generated melt pockets predicted by Petricca et al. (2025). If so, such nutrient-rich water pockets – putatively capable of sustaining heterotrophic biospheres – could thus erupt onto the surface through effusive cryovolcanism. If upheld, this possibility would further raise the astrobiological interest of Titan's cryovolcanic regions.

Besides a putative access to Titan's interior, effusive cryovolcanism offers a direct pathway for organic–water interactions on Titan's surface: erupted cryomagma flows would immediately encounter the thick blanket covering much of the landscape. Helpfully, the surface pressure on Titan (~1.5 bar) would suppress quick vaporization of most aqueous materials (Morrison et al. 2022), prolonging flow runout and extending the window of contact between liquid water and the organic sediments.

A second indication of cryovolcanic activity has come from the north polar region, where a distinctive class of small depressions termed rampart craters has been attributed to explosive cryovolcanism (**Figure 2**). Rampart craters average ~10 km in diameter and possess sharp edges, raised rims, and surrounding deposits interpreted as ejecta ramparts (Mitri et al. 2019; Wood and Radebaugh 2020; Brouwer et al. 2024). Mitri et al. (2019) modelled their formation after the vaporization of liquid nitrogen in the near-subsurface, induced by a warming event. Wood and Radebaugh (2020) proposed an alternative scenario in which a relatively warm aqueous cryomagma rising from depth encounters and vaporizes nitrogen or methane contained in a near-surface alkanofer (i.e., a subsurface reservoir of liquid hydrocarbons), driving a maar-like explosion. More recently, Brouwer et al. (2024) introduced an additional formation hypothesis inspired by gas emission craters found in terrestrial permafrost regions, where methane released from destabilizing clathrates accumulates under an impermeable cap and pressurizes before exploding. Similarly to earlier proposals, the latter process would require a subsurface thermal anomaly to initiate clathrate breakdown (Brouwer et al. 2024).

Explosive cryovolcanism would enable only limited interaction between liquid water and surface organic material. The explosive excavation of material would fragment and eject both subsurface ice and any entrained organic sediments, depositing them as a mixed blanket (Wood and Radebaugh 2020; Brouwer et al. 2024). Any aqueous component dispersed explosively would freeze rapidly, precluding prolonged liquid water–organic interactions. Still, some raised ramparts show flow-like morphologies suggesting that cryomagma could have also erupted in these craters (Wood and Radebaugh 2020).

The third manifestation of Titan's cryovolcanism is intrusive cryomagmatism. Because aqueous cryomagmas are predominantly negatively buoyant with respect to the ice shell, intrusive processes (i.e., the subsurface emplacement of liquid water or ammonia–water mixtures) may greatly dominate over extrusive eruptions, by analogy with silicate volcanism on Earth, Mars, and Venus (Schurmeier et al. 2023). Schurmeier et al. (2023) proposed that the radially networked labyrinth terrains found in Titan's northern mid-latitudes represent the surface expression of deep cryomagmatic laccoliths – large volumes of liquid water emplaced as intrusions within the ice shell. These intrusions would cool on timescales far longer than surface flows, potentially providing long-lived environments nurturing aqueous organic chemistry. The latter would be fed by the same delivery pathways discussed for the interior hydrosphere (section 2), likely complemented by the percolation of organic-laden fluids through fractures and pore networks in the overlying ice (Hayes et al. 2008, 2017; Malaska et al. 2022). The extent of such percolation remains, however, unclear.

### 4.2. Prebiotic chemistry in cryovolcanic melts

Any interaction between cryovolcanic aqueous melts and organic molecules is very likely to yield prebiotic molecules. Indeed, a large body of laboratory work has demonstrated that Titan's organic macromolecules undergo quick hydrolysis across a range of pH conditions: in acidic solutions at elevated temperatures (Khare et al. 1986; Taniuchi et al. 2013), in pure water at neutral pH (Neish et al. 2008), and – most relevant for the ammonia–water model of Mitri et al. (2008) – in alkaline ammonia–water solutions (Neish et al. 2009, 2010; Ramírez et al. 2010; Poch et al. 2012; Cleaves et al. 2014; Brassé et al. 2017; Farnsworth et al. 2024). The experiments of Neish et al. (2009) initially showed that complex tholin macromolecules placed in 13 wt% ammonia–water incorporate oxygen from the water molecules into their molecular structure; over the course of one year, tholins hydrolyzed in this solution incorporated 6% oxygen by mass at 253 K and 12% at 293 K, with four amino acids identified among the products (Neish et al. 2010). Ramírez et al. (2010) also demonstrated that a large diversity of amino acids remains detectable when hydrolyzing tholins in 3–25 wt% ammonia at temperatures as low as 96 K, though yields decrease strongly at lower temperature. Poch et al. (2012) confirmed these trends using 25 wt% ammonia and identified urea as the dominant hydrolysis product, with glycine and alanine being also detected. Cleaves et al. (2014) extended the experimental timescale to 2.5 years (14 wt% ammonia, 25°C), detecting glycine as the most abundant amino acid alongside alanine and γ-aminobutyric acid. Brassé et al. (2017)

combined a lower ammonia concentration of 5 wt% with carbonates to yield urea, again, as the principal product, alongside several amino acids, carboxylic acids, and fatty acids relevant to prebiotic membrane chemistry. More recently, Farnsworth et al. (2024) investigated the alkaline hydrolysis of aminonitriles, putative intermediates in the hydrolysis of photochemical material into amino acids. At 15 wt% $NH_3$ and −22°C they measured alanine yields of 3.3%, after six months, from its aminonitrile precursor, demonstrating that the aminonitrile pathway is a productive channel for amino acid synthesis, at temperatures close to the melting point of ammonia–water cryomagmas (Farnsworth et al. 2024).

Nevertheless, the principal limitation on the prebiotic potential of cryovolcanic regions has been the competition between reaction kinetics and the cryomagma freezing timescale (Neish et al. 2018). Extrapolating the measured Arrhenius activation energies from ~253 K down to the ammonia–water peritectic at 176 K yields half-lives of hundreds of years for most observed hydrolysis reactions (Neish et al. 2009; Lunine et al. 2020). Yet effusive flows can freeze in ~1 year (Neish et al. 2006, 2018; Lunine et al. 2020). The kilometer-scale intrusive cryomagma bodies proposed by Schurmeier et al. (2023) would extend the window available for aqueous prebiotic chemistry by orders of magnitude relative to surface flows, though the limited delivery of surface organics to such depths would limit reaction rates.

### 4.3. Detection by Dragonfly and future missions

Dragonfly is not expected to fully establish the prebiotic potential, properties, or existence of Titan's cryovolcanism. That is chiefly because of a geographical mismatch between the equatorial regions accessible to Dragonfly – landing at ~4°N – and the high-latitude locations where cryovolcanism seems most predominant. Although the strongest candidate for extrusive cryovolcanism – the Sotra Patera complex – sits at ~15°S, the radial labyrinth terrains (intrusive cryomagmatism) are found at mid-latitudes (~39°–42°N; Malaska et al. 2020) and the rampart craters (explosive cryovolcanism) at the north polar region. This apparent concentration of cryovolcanic features toward the poles may reflect the lower elevation of the polar regions relative to the equator (Zebker et al. 2009; Corlies et al. 2017), and their likely thinner and warmer ice crust, conditions that would favor cryomagma ascent (Durante et al. 2019; Wood and Radebaugh 2020).

Gonçalves et al. (2025b) proposed a mechanism by which prebiotic molecules produced in the polar cryovolcanic regions could reach the equatorial latitudes sampled by Dragonfly. The authors proposed that the exsolution of gases dissolved in the cryomagmas would thrust into the atmosphere ammonia–water aerosols encapsulating the cryomagmas' organic components. Once made airborne, the ammonia–water icy aerosols would be transported equatorward by Titan's Hadley circulation (Mitchell and Lora 2016; Lombardo and Lora 2023), permitting it to deposit onto equatorial regions (Gonçalves et al. 2025b). A complementary pathway invokes Titan's high-latitude hydrocarbon seas: prebiotic molecules produced in cryovolcanic regions could be transported by methane–ethane runoff into the polar lakes and seas, where bubble bursting from nitrogen-supersaturated liquids would aerosolize them into droplets (Cordier et al. 2024). These could then be subjected to the same meridional redistribution. Consistent with that transport scenario, fractures and faults formed by tectonic stress can host fluvial networks (Burr et al. 2013); hydrocarbon fluvial flow along such tectonic structures should thus expose fresh material, favoring the leaching of organics associated with tectonically induced cryovolcanism. Through these transport mechanisms, Dragonfly may sample cryovolcanogenic amino acids sedimented in the equatorial dune fields, despite their distance to the cryovolcanic features. The amount of material thus transported should be low, but perhaps sufficient for detection by DraMS, which is designed to identify amino acids and other prebiotic molecules at parts-per-billion sensitivity (Barnes et al. 2021; Moulay et al. 2023; Boulesteix et al. 2024).

The Dragonfly mission will also produce high-resolution imagery of multiple regions along its traverse that will place cryovolcanism in context with fluvial, aeolian, and impact processes (Barnes et al. 2021). Still, significant progress on the question of cryovolcanism will require dedicated orbital or aerial platforms capable of high-resolution imaging (e.g., Coustenis et al. 2009; Mitri et al. 2014a; Rodriguez et al. 2022), particularly over the polar terrains (Nixon et al. 2025). Landers dedicated to the exploration of the Sotra Patera complex (Neish et al. 2018) or flybys performed by future missions to the Saturnian system – such as ESA's upcoming L4 mission (Martins et al. 2024) – could also break through the limitations of Cassini's data.

Furthermore, the mechanics of cryovolcanism remain poorly constrained owing to insufficient knowledge of the thickness, composition, and activity of the outer ice shell (MacKenzie et al. 2021). The diversity of proposed mechanisms may reflect a representative variety or merely uncertainty from lack of constraints. Future laboratory work simulating the buoyancy, eruption, and freezing of cryomagmas, with compositions encompassing eutectic ammonia–water and carbonates and other salts, would meaningfully reduce these uncertainties (Zhong et al. 2009; Dougherty et al. 2018; Munoz-Iglesias and Prieto-Ballesteros 2022; Morrison et al. 2022). Until they are resolved, the prebiotic potential of Titan's cryovolcanic environments, though compelling, remains an open question.

## 5. Impact Craters

Impacts into Titan's icy crust generate transient pools of liquid water intimately mixed with photochemically derived organics and exogenous material delivered by the impactor (**Figure 1c**). These conditions are highly favorable to aqueous organic chemistry and significantly more permissive than those of cryovolcanic melts. Whereas the latter should be limited to temperatures near the ammonia–water peritectic (~176 K), impact melts provide ample thermal energy for hydrolysis and related reactions to proceed (Lunine et al. 2020; Hedgepeth et al. 2022), even approaching chemical equilibrium (Madan and Pearce 2025). The melt environments are also long-lived: their refreezing timescales range from hundreds to tens of thousands of years, depending on impactor size and melt composition (Artemieva and Lunine 2003; O'Brien et al. 2005; Hedgepeth et al. 2022; Kalousová et al. 2024; MacKenzie et al. 2024). The combination of extensive water–organic interaction, elevated temperatures, and extended durations makes impact craters the most promising surface environments on Titan for prebiotic synthesis (Neish et al. 2018; MacKenzie et al.

2024; section 5.1) and motivates the selection of Selk crater as a primary target of Dragonfly (Barnes et al. 2021; section 5.2).

Titan's thick atmosphere shields the surface from small impactors, a filtering effect that accounts in part for the relatively low number of confirmed impact craters identified on the surface (Wood et al. 2010; Neish and Lorenz 2012). The most recent comprehensive survey of Cassini data identified approximately 90 impact structures (Hedgepeth et al. 2020). This modest inventory also reflects the low crater retention age (Neish and Lorenz 2012; Wakita et al. 2026), efficient crater relaxation (Schurmeier and Dombard 2018; Schurmeier et al. 2024; Brouwer et al. 2025) and degradation by resurfacing processes like fluvial erosion (Neish et al. 2013, 2016), sediment transport (Lopes et al. 2020; Lapôtre et al. 2022), mechanical weathering (Seltzer et al. 2025), and burial (Shah et al. 2025). Despite the limited number of recognized structures, the total estimated melt production across Titan is on the order of ~600,000 $km^3$ (MacKenzie et al. 2024), a substantial cumulative volume of aqueous chemical reactors.

The distribution of these impact craters is also notably non-uniform (cf. Wakita et al. 2022). Most confirmed impact structures are concentrated in the equatorial and mid-latitude regions while both polar regions show a marked deficit: 65% of Titan's craters are found within ±30° of the equator (Hedgepeth et al. 2020). Interestingly, the predominance of impact craters at low latitudes favors the interaction between impact melts and organic sediments, themselves also mostly accumulated in the equator (Solomonidou et al. 2018; Lopes et al. 2020). Spectral analysis of nine craters by Solomonidou et al. (2020) reinforced this picture: equatorial dune craters – including Dragonfly's Selk – are spectrally dominated by organic-rich material with no discernible water-ice signature. If such organic infilling began while the impact melts were still liquid, Titan's equatorial impact craters would have interacted directly with a plethora of organic material, and very likely harbored extensive prebiotic chemistry.

### 5.1. Prebiotic chemistry in impact-melt pools

As noted above, Titan's surface receives a sustained flux of small organic molecules settling out of the atmosphere. These constitute a standing reservoir of feedstocks available for aqueous processing wherever liquid water is transiently produced. In impact craters, this reservoir is supplemented by organic molecules delivered by the impactor itself, including molecules of direct prebiotic relevance (e.g., Hadraoui et al. 2019). Beyond increasing the molecular inventory, impactors also introduce exogenic chemical elements absent or scarce in Titan's atmospheric chemistry – most notably phosphorus and sulfur (Pasek et al. 2011; Martins and Pasek 2024; MacKenzie et al. 2024) – opening reaction pathways otherwise unavailable on the surface. For instance, exogeneous phosphorus is expected to exist on Titan's impact melts in the form of phosphates, reduced oxides (namely phosphites and hypophosphites), and phosphonates, which can all act as activated phosphorus compounds towards the formation of phosphate biomolecules (Pasek et al. 2011). Taken together, these contributions can yield prebiotic molecule concentrations in impact melt pools orders of magnitude higher than those attainable by hydrothermal synthesis alone (MacKenzie et al. 2024).

A central question is how much of this endogenous and exogenous organic inventory survives the shock and heating of impact. This was recently quantified by Pearce et al. (2024a), who subjected haze analogues to shock pressures representative of the average impact velocity on Titan's icy surface and found that overall biomolecule concentrations were reduced by factors of 3–8. Still, the authors also noted that the initial shock would also modify the pre-existing organic inventory in prebiotically productive ways, consistent with Martins et al. (2013). Certain species – including thymine, isoleucine/norleucine, and aspartic acid – were found to be enriched or newly synthesized under shock conditions, a production that also appears to be enhanced by mineral surface catalysis in the presence of quartz sand (Pearce et al. 2024a). Further work is required on the immediate consequences of impacts under Titan-specific surface conditions, particularly involving water-ice targets at cryogenic temperatures. Nevertheless, the available evidence supports the view that impact craters are sites of accumulation and exogenous delivery where rich pathways of shock chemistry occur.

Following the dissipation of the shock energy, chemical complexification is very likely to proceed within the melt pool. Neish et al. (2008) demonstrated that complex organic macromolecules synthesized under Titan-like atmospheric conditions undergo hydrolysis when placed in liquid water, incorporating oxygen at timescales far shorter than the lifetimes of the impact melt pools. Farnsworth et al. (2024) further explored this chemistry by experimentally simulating the hydrolysis of aminonitriles under warm impact melt conditions (liquid water at 3°C and 21°C), having demonstrated the production of glycine, alanine and γ-aminobutyric acid. Recent thermodynamic modeling by Madan and Pearce (2025) also identified an energetically favorable route to alanine synthesis requiring only hydrogen cyanide, acetylene, and water, with acrylonitrile ($CH_2CHCN$) as intermediate. In favor of such prebiotic syntheses, their estimated equilibration times – 1–123 years (Madan and Pearce 2025) – are far shorter than the melt pool lifetimes predicted for Selk-size craters – $10^3$–$10^4$ years (Wakita et al. 2023). Reaction rates also increase as progressive refreezing concentrates the solutes (Hedgepeth et al. 2022; MacKenzie et al. 2024).

Impact craters offer favorable conditions for prebiotic chemistry, but its efficiency depends strongly on ammonia. Ammonia is expected not only to accelerate hydrolysis – through the elevation of pH – toward biologically relevant end-products (Neish et al. 2010) but also to open reaction pathways unavailable in pure water (Neish et al. 2009). Already hinted at by previous experimental works (Poch et al. 2012), the interpretation of an enhanced hydrolysis rate driven by heightened pH was demonstrated by Farnsworth et al. (2024). Upon negligible production of amino acids, over six months, in pure water solutions, the addition of at least 5% $NH_3$ increased the yield of glycine, alanine and γ-aminobutyric acid by 2–3 orders of magnitude (Farnsworth et al. 2024). These observations were complemented by the thermodynamic modeling of Madan and Pearce (2025), which showed that the introduction of ammonia expands the set of thermodynamically accessible amino acids from a narrow trio (alanine, β-alanine, proline) to nearly the full canonical suite of proteinogenic amino acids. Subsequent work concluded that biomolecules from other classes (nucleobases, sugars, and fatty acids) become thermodynamically accessible solely in the presence of ammonia. Otherwise, only alanine, β-alanine, proline, adenine, and butanoic acid are thermodynamically favorable (Madan and Pearce 2026). The expected scarcity of ammonia at impact sites may therefore significantly limit their prebiotic potential.

Although the presence of ammonia in impact craters has been frequently considered a possibility – via grain-boundary inclusions (Nasralla et al. 2025) or through destabilization of methane clathrate during impact (Choukroun and Sotin 2012) – a more plausible scenario is that the bulk outer ice shell is composed predominantly of pure water ice (possibly with methane clathrate in the upper portion), with ammonia excluded by the phase equilibria of top-down freezing. Consistent with this assessment, VIMS data show no indication of ammonia ice in Titan's impact craters, even in those less covered by sand sediment and whose ice material is more exposed to VIMS measurements (Solomonidou et al. 2020). The ammonia concentration in these environments will be resolved by the upcoming molecular analysis from DraMS and bulk elemental measurements from DraGNS at Selk crater (Barnes et al. 2021). The results of these measurements will carry significant astrobiological implications.

### 5.2. Sampling Selk crater with Dragonfly

Selk crater embodies the features that make Titan's impact craters compelling targets for prebiotic chemistry investigations. Situated at 7°N, 199°W, near sand seas, this ~84 km crater is among Titan's larger and fresher craters; its depth-to-diameter ratio indicates an intermediate degradation state, preserving much of the original impact structure while remaining accessible to surface exploration (Neish et al. 2018; Hedgepeth et al. 2020). Numerical simulations by Wakita et al. (2023) show that the crater's dimensions are reproduced by a ~4 km diameter icy impactor striking a surface underlain by a 10–15 km thick methane clathrate layer, generating approximately 360–400 $km^3$ of impact melt – a volume substantially larger than earlier estimates based on pure water-ice targets (Artemieva and Lunine 2003). The methane clathrate cap would also extend the melt lifetime relative to a pure water-ice target, to tens of thousands of years (Wakita et al. 2023; Kalousová et al. 2024). This would be enough time for detectable prebiotic chemistry to have occurred, largely fed by the organic-rich dune sands that extend beyond its rim (Soderblom et al. 2010; Solomonidou et al. 2020).

Still, the chemical characterization of Selk's frozen melt pool faces a practical challenge: the most chemically evolved material is concentrated at depth. Modeling by Hedgepeth et al. (2022) demonstrates that as the melt pool freezes downward, dissolved organics are continuously rejected into the residual liquid, amplifying their concentration progressively until they reach a maximum at approximately 75% into the melt sheet relative to the surface. This is the locus of the longest-lived liquid water and therefore the site of the most complete aqueous chemistry. Accessing this zone directly would require drilling capabilities beyond Dragonfly's scope. Neish et al. (2018) noted that Titan's geomorphic processes may provide a natural solution: methane rivers are known to incise into Titan's impact craters (Neish et al. 2016, 2018; Hedgepeth et al. 2020), cutting through the frozen melt sheet and exposing its interior along riverbed walls, while simultaneously depositing eroded material in downstream alluvial fans (Neish et al. 2018). Both features – exposed riverbed walls and alluvial fans – can allow direct access to the formerly liquid interior of the melt deposit. Encouragingly, the upper 10–50 m reachable by such erosional processes would have solidified on timescales of approximately 0.5–10 years (Hedgepeth et al. 2022; Madan and Pearce 2025); although far shorter than the deep melt lifetime, these timescales are still large enough for at least partial chemical equilibration (Madan and Pearce 2025), and thus for representative synthesis of prebiotic molecules. Furthermore, prebiotic molecules produced within these frozen melt deposits are expected to be exceptionally well preserved by Titan's dense atmosphere. The atmosphere not only attenuates high-energy irradiation at the surface but shields it from small-impactor gardening, meaning that aqueous organic products generated within the melt – even if scarce – have likely been preserved (Lunine et al. 2020).

Dragonfly's investigation of Selk will be the culmination of a traverse that begins at the Shangri-La dune field (**Figure 3**). Shangri-La comprises dunes hundreds of kilometers long that accumulate organic sediments that have never interacted with liquid water (Lorenz et al. 2021). By comparing the molecular inventory of these pristine dune sands with that of the impact melt at Selk, DraMS will be able to identify the compositional changes driven by the impact event; specifically, it will determine which organic precursors were transformed into amino acids and other biomolecules in the transient liquid water environment (Barnes et al. 2021). Complementing the organic inventory, DraGNS will measure bulk elemental abundances at each landing site, enabling the search for elements scarce on Titan but essential for habitability (e.g., sulfur and phosphorus); enrichment of these elements in impact-related terrains would demonstrate that impacts serve as viable delivery mechanisms (MacKenzie et al. 2024). DragonCam may further constrain the degree of post-impact erosion at Selk, testing predictions of fluvial incision and aeolian infill in the crater morphology (Neish et al. 2016; Hedgepeth et al. 2020) and informing the extent of water–organic interaction in these settings.

In between the dune region and the crater floor, Dragonfly will also sample Selk's ejecta blanket (**Figure 3**; Barnes et al. 2021). Cassini RADAR radiometry at 2.18 cm – which probes to greater depths than VIMS – suggests a water-ice-enriched subsurface in the ejecta blanket (Werynski et al. 2019), whereas VIMS spectra of the same region show that the uppermost surface layer is dominated by organic material with no detectable water-ice signature (Solomonidou et al. 2020). This apparent contrast between the near-infrared surface composition and the microwave subsurface signature is consistent with a scenario in which the impact excavated water and mixed it with the ambient organic inventory; the resulting deposit was subsequently covered by organic sediments accumulated over geological timescales (Werynski et al. 2019; Solomonidou et al. 2020). The subsurface water–organic mixture, cooled away from the warm central melt pool, will provide a comparison between distinct thermal processing regimes. Another notable landform within the ejecta is the hundred-kilometer-long feature located east-southeast of Selk that may represent a fluidized ejecta flow. Designated the "bench", it potentially preserves impact melt that interacted extensively with the organic-rich sands (Soderblom et al. 2010). Overall, if Dragonfly's measurements indicate that aqueous processing also occurred within the ejecta deposits – not solely in the central melt sheet – the spatial extent of prebiotically productive aqueous environments associated with a single impact event would expand correspondingly, amplifying the prebiotic potential of Titan's crater population.

## 6. Implications for other environments

Assessing the habitability and prebiotic potential of Titan's aqueous settings (summarized in **Table 1**) provides a benchmark for other worlds. Impact cratering is the clearest case in point. As advanced by MacKenzie et al. (2024), impact craters across ocean worlds can be thought of as natural laboratories in which an impact event initiates an aqueous chemical experiment whose timescales – most notably Titan's – far exceed those achievable in the laboratory. Accordingly, Titan's Selk crater will provide a vital data point in understanding the overall prebiotic potential of impact craters. The same extrapolation holds for studies of Titan's cryovolcanism. These may inform, for instance, the interpretation of Pluto's red Virgil Fossae deposit, putatively emplaced as a tholin-laden ammonia–water cryomagma sourced from a subsurface reservoir (Neveu et al. 2015; Cruikshank et al. 2019). By analogy with Titan tholin hydrolysis experiments, this deposit is expected to harbor a broad inventory of prebiotic molecules (Cruikshank et al. 2019). More broadly, Titan's aqueous settings constitute natural experiments in the transition from prebiotic to hypothetical biotic chemistry. Characterizing them allows us to map the space of abiotic possibilities against which putative "biosignature thresholds" can be defined (Barge et al. 2022; MacKenzie et al. 2024). This framing elevates Dragonfly into a mission probing not only for potential biosignatures but also for the boundaries of the abiotic complexity arising in organic-rich aqueous environments (Barnes et al. 2021; Madan and Pearce 2026).

Titan also provides a powerful bridge between Solar System habitability studies and exoplanet science. It serves as a calibration laboratory in which aerosol formation, growth, vertical distribution, and surface deposition can be studied through both remote observations and *in situ* surface measurements. The study of these processes – and the consequences for their habitability – are unavailable at any other hazy world. Titan studies help calibrate the atmospheric retrieval frameworks needed for the detectability and interpretation of the chemistry of new classes of exoplanets now being proposed – including "warm exo-Titans" (e.g., Ranjan et al. 2025) –, with James Webb Space Telescope (JWST) observations already deepening analogies between Titan and candidate Titan-like exoplanets (e.g., Mandt et al. 2022; Glidden et al. 2025). By anchoring aerosol microphysics and optical properties to Titan observations, we can substantially improve the interpretation of muted or featureless exoplanet spectra (Niraula et al. 2025), particularly for planets orbiting M-dwarf stars, where photochemistry is expected to be influenced by frequent UV flaring (e.g., Lunine 2009). This is directly relevant for upcoming facilities such as the Habitable Worlds Observatory (HWO), aimed at characterizing Earth-sized habitable planets across a range of atmospheric conditions. The synergy between Titan exploration and exoplanet-dedicated observatories is particularly important for ESA's upcoming Atmospheric Remote-sensing Infrared Exoplanet Large-survey (ARIEL) mission, which will conduct the first large-scale spectroscopic survey of exoplanet atmospheres, characterizing the chemical composition of the atmosphere, clouds, and hazes of more than 1,000 planets (Tinetti et al. 2018; Chubb et al. 2024). Exploring that synergy, Titan observations with the ground-based Very Large Telescope-Echelle SPectrograph for Rocky Exoplanet and Stable Spectroscopic Observations (VLT-ESPRESSO) have enabled detailed characterizations of methane absorption features in Titan's visible spectrum (Rianço-Silva et al.

2024; Sithajan et al. 2025) and refined the search for trace atmospheric constituents (e.g., Rianço-Silva et al. 2026), overall offering novel constraints on atmospheric photochemical pathways and their detection. Further future observatories such as the Extremely Large Telescope (ELT), HWO, and the Large Interferometer For Exoplanets (LIFE) will enable direct imaging and reflectance spectroscopy of non-transiting worlds, while targeting exomoons, particularly in M-star systems (e.g., Martínez-Rodríguez et al. 2019). Titan is due to serve as benchmark for both.

Titan is, finally, central to understanding how atmospheres form and evolve. In particular, it is the closest known analogue to Earth's Archean hazy atmosphere (Arney et al. 2016), with its methane-driven atmospheric organic production providing a model both for early-Earth haze formation (Trainer et al. 2006; Hörst et al. 2018) and for how photochemistry may have contributed to the inventory of prebiotic molecules on the early Earth's surface (He and Smith 2014; Pearce et al. 2024b). Furthermore, Titan informs one of the central open questions in exoplanet science: why only some rocky worlds retain thick atmospheres. The contrast between Titan and the Jovian moon Ganymede – similar in size yet essentially airless – highlights the importance of formation pathways, volatile inventories, and ice-line chemistry in shaping atmospheric retention. This dichotomy is relevant for characterizing some exoplanetary systems, with models predicting that Titan-like worlds may exist at the outer edges of compact systems such as TRAPPIST-1 (e.g., Mandt et al. 2022; Glidden et al. 2025). Titan's size, icy–rocky composition, and outer-Solar-System location provide a critical data point. By studying its habitability, we gain a framework for interpreting habitability across a broad diversity of worlds.

## 7. Conclusion

All three aqueous settings reviewed here – Titan's interior, its cryovolcanic regions, and its impact craters – permit the co-occurrence of liquid water, bioessential elements, and energy sources. Each does so through markedly different mechanisms, and each faces distinct bottlenecks to its prebiotic potential. Yet all satisfy the standard criteria for habitability (Cockell et al. 2016).

Their sustenance is a harder question. The longer habitable conditions are sustained, the more likely they are to enable the chemical evolution leading to life (Cockell et al. 2024), but current evidence suggests Titan's aqueous environments are short-lived on geological timescales. Impact-melt pools refreeze on $10^2$–$10^4$ yr timescales (Neish et al. 2018; Hedgepeth et al. 2022; Wakita et al. 2023), and putative cryovolcanic flows have shorter lifetimes still. The one environment with geological longevity would be the subsurface ocean: under the standard interior model, it would have been liquid for ~4.5 Gyr (Lunine et al. 2020). During that time, it could have been continuously replenished with material released from the core, thus sustaining the chemical disequilibria and the nutrient supply needed for chemical complexification and, putatively, a sustained heterotrophic biosphere (Miller et al. 2026). However, that best-case scenario is no longer secure. Under the revised interior model (Petricca et al. 2025), the liquid water phase may be transient, tied to Titan's currently elevated orbital eccentricity, which is expected to damp on ~30 Myr timescales (Downey and Nimmo 2025). Given how consequential this distinction is for Titan's sustained habitability (Cockell et al. 2024), the constraints Dragonfly will return on Titan's interior will provide critical ground-truth data.

With less than a decade until Dragonfly's arrival, the coming years offer a closing window to sharpen predictions, prepare interpretive frameworks, and identify the measurements that will most powerfully discriminate between competing scenarios. Three priorities, in particular, would maximize the astrobiological return from this landmark mission: better constraining the processes by which interior material erupts onto the surface; characterizing the transport mechanisms that could deliver cryovolcanogenic prebiotic molecules within Dragonfly's reach, along with their compositional signatures; and developing Titan-specific impact experiments to inform the prebiotic chemistry expected at Selk crater. Titan is where the laboratory chemistry of the past five decades meets a planetary surface; Dragonfly is how we will measure its predictive power.

## 8. Acknowledgements

The authors acknowledge funding by the Portuguese Foundation for Science and Technology (FCT) through fellowship 2021.04932.BD and project ORIGINS (2022.05284.PTDC). This work was financially supported by IMS LA/P/0056/2020 (https://doi.org/10.54499/LA/P/0056/2020), CQE UID/00100/2025 (https://doi.org/10.54499/UID/00100/2025), UID/PRR/100/2025 (https://doi.org/10.54499/UID/PRR/00100/2025) and UID/PRR2/00100/2025 (https://doi.org/10.54499/UID/PRR2/00100/2025) funded by national funds through FCT/MECI (PIDDAC). L.F.C. is funded by the Heising-Simons Foundation 51 Pegasi b postdoctoral fellowship 2024-5174.

**Table 1**. Comparative assessment of the three aqueous habitats reviewed here – the subsurface hydrosphere (section 2), cryovolcanic melts (section 4), and impact-melt pools (section 5) – listing the principal factors enabling habitability and the bottlenecks that limit their prebiotic potential.

| Aqueous habitat | Enabling factors | Main bottlenecks |
|---|---|---|
| Subsurface hydrosphere | • Vast, persistent liquid water – either as a global ocean (Iess et al. 2012) or as melt pockets (Petricca et al. 2025).<br>• Organic building blocks and energy sources delivered from the carbonaceous core (Néri et al. 2020; Reynard and Sotin 2023; Miller et al. 2026) and from the surface via impacts and melt foundering (Crósta et al. 2021; Neish et al. 2024; Kalousová et al. 2024). | • Under a global ocean, the organic input is severely diluted; surface-delivered organics cannot sustain a heterotrophic biosphere via glycine fermentation (Neish et al. 2024; Affholder et al. 2025); whether core-derived organics can do so remains an open question (Miller et al. 2026).<br>• Under the mushy hydrosphere of Petricca et al. (2025), the melt-bearing state has a short lifetime. |
| Cryovolcanic melts | • Direct access to the organic-rich surface reservoir (Janssen et al. 2016; Lopes et al. 2020; Solomonidou et al. 2020; Neish et al. 2024), which provides both prebiotic feedstocks and chemical energy sources (Niemann et al. 2010; Lunine et al. 2020; Das et al. 2025).<br>• Effusive flows place ammonia–water melts in immediate contact with the surface organic blanket (Lopes et al. 2013).<br>• Intrusive cryomagmatic bodies may remain liquid over geological timescales (Schurmeier et al. 2023). | • The very existence of present-day cryovolcanism remains observationally unsettled (Nixon et al. 2025).<br>• Effusive surface manifestations are short-lived and should operate near the ammonia–water peritectic (~176 K); at these temperatures, hydrolysis rates will be slow relative to refreezing timescales, limiting prebiotic yields (Neish et al. 2018; Lunine et al. 2020).<br>• Intrusive cryomagmas are likely depleted in organic material. |
| Impact-melt pools | • Direct contact between water melt pools and the atmospheric organic reservoir.<br><br>• Exogenous delivery by the impactor of additional organics and chemical elements (MacKenzie et al. 2024).<br><br>• Warm and long-lived melt pools (Wakita et al. 2023; Kalousová et al. 2024) sufficient for prebiotic syntheses to approach equilibrium (Farnsworth et al. 2024; Madan and Pearce 2025). | • The expected scarcity of ammonia at impact sites (Solomonidou et al. 2020) significantly restricts both the formation kinetics (Neish et al. 2009, 2010; Farnsworth et al. 2024) and the thermodynamic accessibility (Madan and Pearce 2025, 2026) of biomolecules. |

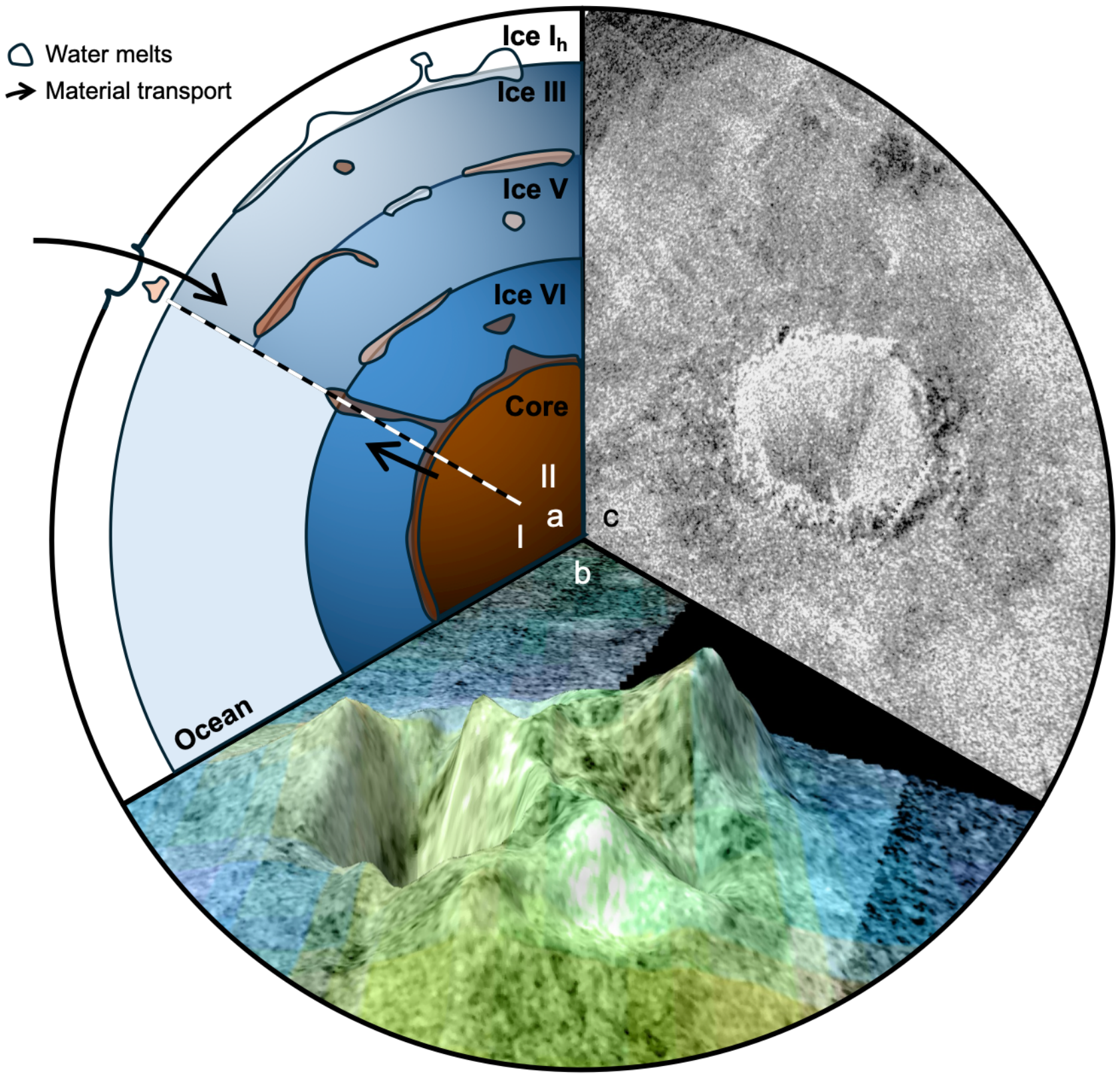

**Figure 1.** Titan's three aqueous settings reviewed here. **(a)** Schematic cross-section of Titan's interior, contrasting two competing models of its hydrosphere (not to scale). The anticlockwise half (I) shows the global-ocean model, in which a deep liquid-water ocean is sandwiched between the outer ice-$I_h$ crust and a layer of high-pressure ice (Sotin et al. 2021). The clockwise half (II) shows the model recently proposed by Petricca et al. (2025), in which the hydrosphere is instead a sequence of high-pressure ice phases (Ice III, V, and VI, in order of increasing depth) interspersed with discrete melt pockets. Arrows indicate the principal pathways by which material may be delivered to the hydrosphere: surface-deposited organics can be transported downward by impacts (Crósta et al. 2021; Neish et al. 2024; Kalousová et al. 2024), while accreted material released from a warm core can percolate upward (Kalousová and Sotin 2018; Miller et al. 2026). Both pathways may enrich the melt pockets in organic material (brown shading). **(b)** Three-dimensional perspective view of the Sotra Patera complex, combining Cassini RADAR-derived topography with a VIMS false-color overlay encoding distinct compositions (NASA image PIA13695). The depicted caldera is ~30 km in diameter (Lopes et al. 2013). **(c)** Cassini SAR image of Sinlap crater, acquired during the T3 flyby (NASA image PIA07368). Sinlap crater is ~90 km in diameter (Hedgepeth et al. 2020). NASA image credits: NASA/JPL-Caltech/ASI/Cassini Radar Team/USGS/University of Arizona.

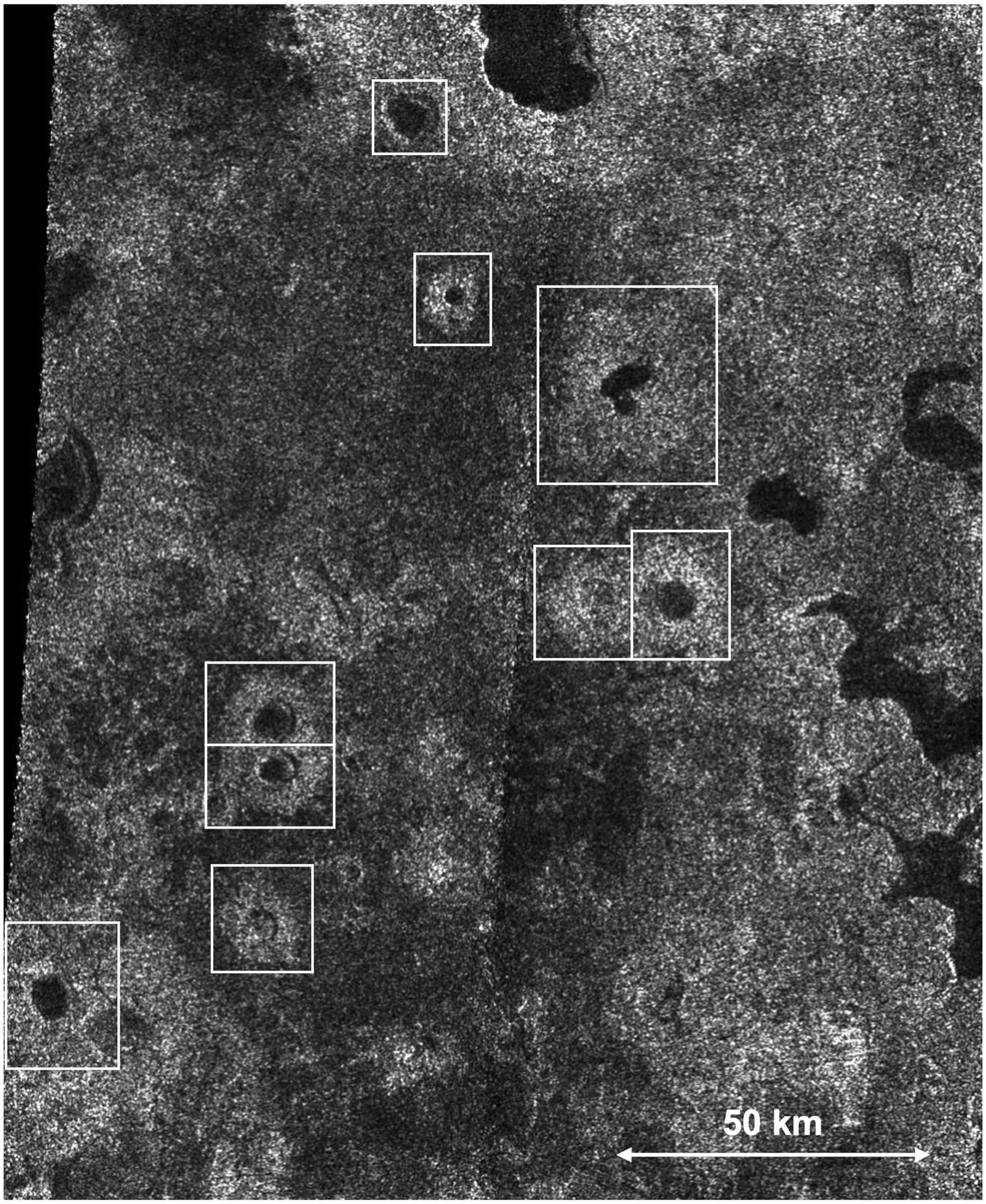

**Figure 2**. Cassini SAR image of a cluster of rampart craters in Titan's north polar lake district, proposed as candidate cryovolcanic features by Mitri et al. (2019), Wood and Radebaugh (2020), and Brouwer et al. (2024). Rampart craters are spatially clustered, generally circular depressions averaging ~10 km in diameter, with raised rims surrounded by radar-bright, positive-relief halos ("ramparts") interpreted as ejecta deposits. Image credit: NASA/JPL-Caltech/ASI/Cassini Radar Team/PDS Imaging Atlas. Crater identification and spatial scale after Brouwer et al. (2024).

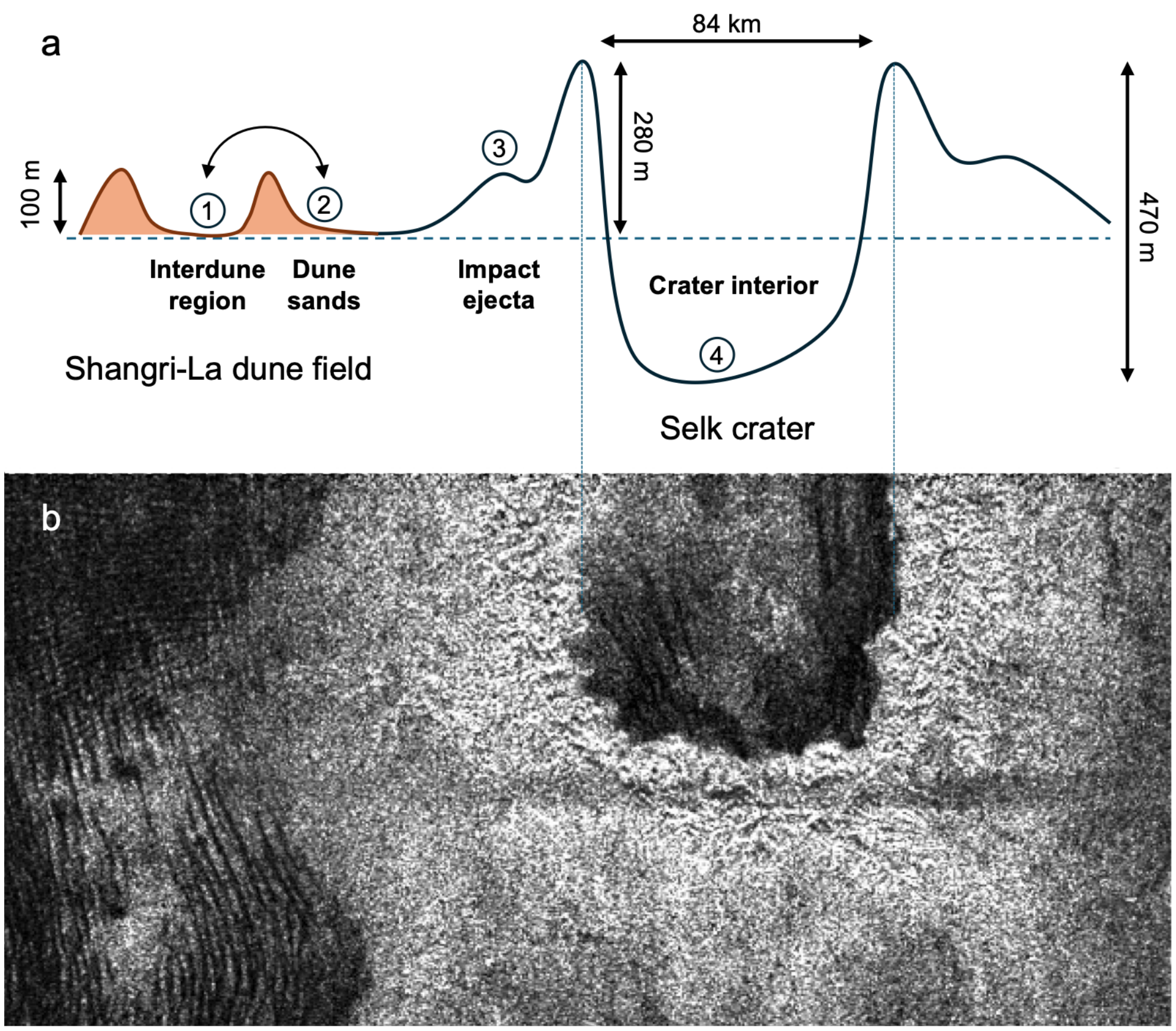

**Figure 3**. Schematic of Dragonfly's traverse and Cassini SAR context for the Selk landing region. **(a)** Idealized topographic profile along the rotorcraft's planned traverse from its initial landing site toward Selk crater. Dragonfly will start by alternating between sand-free interdune (1) and sand dune (2) landing sites. The traverse will then cross the ejecta blanket (3) and rim of Selk before reaching the crater floor (4). The Selk crater dimensions adopted here – diameter: 84 (± 2) km; rim-to-floor depth: 470 (+125/−105) m; rim height: 280 (+50/−54) m – are taken from Hedgepeth et al. (2020). Schematic adapted from Barnes et al. (2021). **(b)** Cassini SAR image of Selk crater and surroundings acquired during the T95 flyby, showing the contrasting textures of the linear dune field (bottom left) and the comparatively radar-bright ejecta blanket and rim of Selk, encircling the radar-dark crater floor (top right) (Lorenz et al. 2021). The two panels are vertically aligned such that the topographic profile in (a) approximately corresponds to a transect through the terrain imaged in (b). Image credit: NASA/JPL-Caltech/ASI/Cassini Radar Team/USGS/PDS Imaging Atlas.

## 9. References

Affholder A, Higgins PM, Cockell CS, et al (2025) The Viability of Glycine Fermentation in Titan's Subsurface Ocean. Planet Sci J 6:86. https://doi.org/10.3847/PSJ/ADBC66

Arney G, Domagal-Goldman SD, Meadows VS, et al (2016) The Pale Orange Dot: The Spectrum and Habitability of Hazy Archean Earth. Astrobiology 16:873–899. https://doi.org/10.1089/AST.2015.1422

Artemieva N, Lunine J (2003) Cratering on Titan: impact melt, ejecta, and the fate of surface organics. Icarus 164:471–480. https://doi.org/10.1016/S0019-1035(03)00148-9

Bains W, Petkowski JJ, Seager S (2024) Alternative Solvents for Life: Framework for Evaluation, Current Status, and Future Research. Astrobiology 24:1231–1256. https://doi.org/10.1089/AST.2024.0004

Baland RM, Tobie G, Lefèvre A, Van Hoolst T (2014) Titan's internal structure inferred from its gravity field, shape, and rotation state. Icarus 237:29–41. https://doi.org/10.1016/J.ICARUS.2014.04.007

Baland RM, Van Hoolst T, Yseboodt M, Karatekin Ö (2011) Titan's obliquity as evidence of a subsurface ocean? Astron Astrophys 530:A141. https://doi.org/10.1051/0004-6361/201116578

Baland RM, Yseboodt M, Van Hoolst T (2016) The obliquity of Enceladus. Icarus 268:12–31. https://doi.org/10.1016/J.ICARUS.2015.11.039

Barge LM, Rodriguez LE, Weber JM, Theiling BP (2022) Determining the "biosignature Threshold" for Life Detection on Biotic, Abiotic, or Prebiotic Worlds. Astrobiology 22:481–493. https://doi.org/10.1089/AST.2021.0079

Barnes JW, Turtle EP, Trainer MG, et al (2021) Science goals and objectives for the Dragonfly Titan rotorcraft relocatable lander. Planetary Science Journal 2:130. https://doi.org/10.3847/PSJ/abfdcf

Bills BG, Nimmo F (2008) Forced obliquity and moments of inertia of Titan. Icarus 196:293–297. https://doi.org/10.1016/J.ICARUS.2008.03.002

Boulesteix D, Buch A, Samson J, et al (2024) Dimethylformamide dimethyl acetal reagent for in situ chiral analyses of organic molecules on Titan with the Dragonfly mass spectrometer space instrument (Dragonfly mission). J Chromatogr A 1722:464860. https://doi.org/10.1016/J.CHROMA.2024.464860

Bradley JA, Arndt S, Amend JP, et al (2020) Widespread energy limitation to life in global subseafloor sediments. Sci Adv 6:. https://doi.org/10.1126/SCIADV.ABA0697

Brassé C, Buch A, Coll P, Raulin F (2017) Low-temperature alkaline pH hydrolysis of oxygen-free Titan tholins: Carbonates' impact. Astrobiology 17:8–26. https://doi.org/10.1089/AST.2016.1524

Brouwer G, Schurmeier L, Fagents S, et al (2025) Topographic Relaxation of Complex Impact Craters in a Clathrate Crust on Titan. Planet Sci J 6:217. https://doi.org/10.3847/PSJ/ADF642

Brouwer GE, Schurmeier LS, Fagents SA (2024) An Endogenic Origin for Titan's Rampart Craters: Assessment of Explosion Mechanisms. J Geophys Res Planets 129:e2024JE008459. https://doi.org/10.1029/2024JE008459

Burks MT, Drury OB, Goldsten JO, et al (2020) GeMini: A High-Resolution, Low-Resource, Gamma-Ray Spectrometer for Planetary Science Applications. Space Sci Rev 216:115-. https://doi.org/10.1007/S11214-020-00746-Y

Burr DM, Drummond SA, Cartwright R, et al (2013) Morphology of fluvial networks on Titan: Evidence for structural control. Icarus 226:742–759. https://doi.org/10.1016/J.ICARUS.2013.06.016

Byrne PK, Regensburger P V, Klimczak C, et al (2018) The Geology of the Rocky Bodies Inside Enceladus, Europa, Titan, and Ganymede. In: 49th Lunar and Planetary Science Conference. p 2905

Cable ML, Hörst SM, Hodyss R, et al (2012) Titan tholins: Simulating Titan organic chemistry in the Cassini-Huygens era. Chem Rev 112:1882–1909. https://doi.org/10.1021/cr200221x

Cable ML, Runčevski T, Maynard-Casely HE, et al (2021) Titan in a Test Tube: Organic Co-crystals and Implications for Titan Mineralogy. Acc Chem Res 54:3050–3059. https://doi.org/10.1021/acs.accounts.1c00250

Carnahan E, Vance SD, Hesse MA, et al (2022) Dynamics of Mixed Clathrate-Ice Shells on Ocean Worlds. Geophys Res Lett 49:e2021GL097602. https://doi.org/10.1029/2021GL097602

Carrasco N, Bourgalais J, Vettier L, et al (2022) A missing link in the nitrogen-rich organic chain on Titan. Astron Astrophys 663:A165. https://doi.org/10.1051/0004-6361/202141025

Castillo-Rogez JC, Lunine JI (2010) Evolution of Titan's rocky core constrained by Cassini observations. Geophys Res Lett 37:L20205. https://doi.org/10.1029/2010GL044398

Chatain A, Le Gall A, Berthelier JJ, et al (2023) Detection and characterization of wind-blown charged sand grains on Titan with the DraGMet/EFIELD experiment on Dragonfly. Icarus 391:115345. https://doi.org/10.1016/J.ICARUS.2022.115345

Choblet G, Tobie G, Sotin C, et al (2017) Heat transport in the high-pressure ice mantle of large icy moons. Icarus 285:252–262. https://doi.org/10.1016/J.ICARUS.2016.12.002

Choukroun M, Grasset O, Tobie G, Sotin C (2010) Stability of methane clathrate hydrates under pressure: Influence on outgassing processes of methane on Titan. Icarus 205:581–593. https://doi.org/10.1016/j.icarus.2009.08.011

Choukroun M, Sotin C (2012) Is Titan's shape caused by its meteorology and carbon cycle? Geophys Res Lett 39:4201. https://doi.org/10.1029/2011GL050747

Chubb KL, Robert S, Sousa-Silva C, et al (2024) Data availability and requirements relevant for the Ariel space mission and other exoplanet atmosphere applications. RAS Techniques and Instruments 3:636–690. https://doi.org/10.1093/RASTI/RZAE039

Clark M, Ravine M, Caplinger M, et al (2025) Optomechanical design of the DragonCam microscopic camera. In: Proceedings SPIE 13599, Optomechanical Engineering 2025. SPIE, pp 71–86

Cleaves HJ, Neish C, Callahan MP, et al (2014) Amino acids generated from hydrated Titan tholins: Comparison with Miller–Urey electric discharge products. Icarus 237:182–189. https://doi.org/10.1016/J.ICARUS.2014.04.042

Cockell CS, Bush T, Bryce C, et al (2016) Habitability: A Review. Astrobiology 16:89–117. https://doi.org/10.1089/AST.2015.1295

Cockell CS, Simons M, Castillo-Rogez J, et al (2024) Sustained and comparative habitability beyond Earth. Nat Astron 8:30–38. https://doi.org/10.1038/s41550-023-02158-8

Coll P, Navarro-González R, Szopa C, et al (2013) Can laboratory tholins mimic the chemistry producing Titan's aerosols? A review in light of ACP experimental results. Planet Space Sci 77:91–103. https://doi.org/10.1016/J.PSS.2012.07.006

Cordier D, Liger-Belair G, Bonhommeau DA, et al (2024) Capillary Processes in Extraterrestrial Contexts. J Geophys Res Planets 129:e2023JE008248. https://doi.org/10.1029/2023JE008248

Corlies P, Hayes AG, Birch SPD, et al (2017) Titan's Topography and Shape at the End of the Cassini Mission. Geophys Res Lett 44:11,754-11,761. https://doi.org/10.1002/2017GL075518

Cottini V, Nixon CA, Jennings DE, et al (2012a) Spatial and temporal variations in Titans surface temperatures from Cassini CIRS observations. Planet Space Sci 60:62–71. https://doi.org/10.1016/j.pss.2011.03.015

Cottini V, Nixon CA, Jennings DE, et al (2012b) Water vapor in Titan's stratosphere from Cassini CIRS far-infrared spectra. Icarus 220:855–862. https://doi.org/10.1016/J.ICARUS.2012.06.014

Coustenis A, Atreya SK, Balint T, et al (2009) TandEM: Titan and Enceladus mission. Exp Astron (Dordr) 23:893–946. https://doi.org/10.1007/S10686-008-9103-Z

Couturier-Tamburelli I, Danger G, Mouzay J, Piétri N (2025) Photochemistry of benzene (C6H6) hydrogen cyanide (HCN) co-condensed ices part 2: Formation of aerosols analogues of titan's atmosphere. Icarus 438:116626. https://doi.org/10.1016/J.ICARUS.2025.116626

Crósta AP, Silber EA, Lopes RMC, et al (2021) Modeling the formation of Menrva impact crater on Titan: Implications for habitability. Icarus 370:114679. https://doi.org/10.1016/j.icarus.2021.114679

Cruikshank DP, Umurhan OM, Beyer RA, et al (2019) Recent cryovolcanism in Virgil Fossae on Pluto. Icarus 330:155–168. https://doi.org/10.1016/J.ICARUS.2019.04.023

Czaplinski EC, Vu TH, Maynard-Casely H, et al (2025) Formation and Stability of the Propionitrile:Acetylene Co-Crystal Under Titan-Relevant Conditions. ACS Earth Space Chem 9:253–264. https://doi.org/10.1021/ACSEARTHSPACECHEM.4C00262

Das K, Gautier T, Szopa C, et al (2025) Reanalysis of the Huygens GCMS dataset - II. Trace species in Titan's lower atmosphere. Astron Astrophys 700:A86. https://doi.org/10.1051/0004-6361/202451466

de Kok R, Irwin PGJ, Teanby NA, et al (2007) Oxygen compounds in Titan's stratosphere as observed by Cassini CIRS. Icarus 186:354–363. https://doi.org/10.1016/J.ICARUS.2006.09.016

Delaroque L, Kawamura T, Lucas A, et al (2026) Investigating the Detectability of Body Wave Phases From Tidal Ice Cracking Events on Titan With the Dragonfly Short-Period Seismometer. J Geophys Res Planets 131:e2025JE009432. https://doi.org/10.1029/2025JE009432

Dougherty AJ, Bartholet ZT, Chumsky RJ, et al (2018) The Liquidus Temperature for Methanol-Water Mixtures at High Pressure and Low Temperature, With Application to Titan. J Geophys Res Planets 123:3080–3087. https://doi.org/10.1029/2018JE005707

Downey BG, Nimmo F (2025) Titan's spin state as a constraint on tidal dissipation. Science Advances 11:4741. https://doi.org/10.1126/SCIADV.ADL4741

Dubois D, Carrasco N, Jovanovic L, et al (2020) Positive ion chemistry in an N2-CH4 plasma discharge: Key precursors to the growth of Titan tholins. Icarus 338:113437. https://doi.org/10.1016/J.ICARUS.2019.113437

Dunaeva AN, Kronrod VA, Kuskov OL (2025) Organic Matter in Titan: Models of Internal Structure. Solar System Research 59:73-. https://doi.org/10.1134/S0038094625600088

Durante D, Hemingway DJ, Racioppa P, et al (2019) Titan's gravity field and interior structure after Cassini. Icarus 326:123–132. https://doi.org/10.1016/J.ICARUS.2019.03.003

Elachi C, Allison MD, Borgarelli L, et al (2005) Radar: The Cassini Titan RADAR mapper. Space Sci Rev 115:71–110. https://doi.org/10.1007/S11214-004-1438-9

Fagents SA, Lopes RMC, Quick LC, Gregg TKP (2022) Cryovolcanism. In: Gregg TKP, Lopes RMC, Fagents SA (eds) Planetary Volcanism across the Solar System. Elsevier, pp 161–234

Farnsworth KK, McLain HL, Chung A, Trainer MG (2024) Understanding Titan's Prebiotic Chemistry: Synthesizing Amino Acids Through Aminonitrile Alkaline Hydrolysis. ACS Earth Space Chem 8:2380–2392. https://doi.org/10.1021/ACSEARTHSPACECHEM.4C00114

Fortes AD (2012) Titan's internal structure and the evolutionary consequences. Planet Space Sci 60:10–17. https://doi.org/10.1016/J.PSS.2011.04.010

Fortes AD, Grindrod PM, Trickett SK, Vočadlo L (2007) Ammonium sulfate on Titan: Possible origin and role in cryovolcanism. Icarus 188:139–153. https://doi.org/10.1016/J.ICARUS.2006.11.002

Francis TA, Cable ML, Maynard-Casely HE, et al (2026) An exploration of the three phenylacetylene polymorphs and discovery of a benzene–phenylacetylene co-crystal by vibrational spectroscopy and powder diffraction. Physical Chemistry Chemical Physics 28:6214–6227. https://doi.org/10.1039/D5CP04929F

Freissinet C, Moulay V, Li X, et al (2024) Unveiling the Nitrogen Chemistry of Titan with the Dragonfly Mass Spectrometer: Experimental Focus on Amines and Amides. ACS Earth Space Chem 8:1832–1846. https://doi.org/10.1021/ACSEARTHSPACECHEM.4C00143

Gautier T, Serigano J, Das K, et al (2024) Reanalysis of the Huygens GCMS dataset - I. High-resolution methane vertical profile in the atmosphere of Titan. Astron Astrophys 690:A165. https://doi.org/10.1051/0004-6361/202244583

Geissler P (2015) Cryovolcanism in the Outer Solar System. In: Sigurdsson H (ed) The Encyclopedia of Volcanoes, 2nd edn. Academic Press, pp 763–776

Glein CR (2015) Noble gases, nitrogen, and methane from the deep interior to the atmosphere of Titan. Icarus 250:570–586. https://doi.org/10.1016/j.icarus.2015.01.001

Glidden A, Ranjan S, Seager S, et al (2025) JWST-TST DREAMS: Secondary Atmosphere Constraints for the Habitable Zone Planet TRAPPIST-1 e. Astrophys J Lett 990:L53. https://doi.org/10.3847/2041-8213/ADF62E

Gonçalves D, Hofmann F, Drauschke J, et al (2025a) Characterization of Amino Acid Nanolayers and Their Interactions under Simulated Planetary Conditions. ACS Earth Space Chem 9:356–368. https://doi.org/10.1021/ACSEARTHSPACECHEM.4C00334

Gonçalves D, Hofmann F, Wipf S, et al (2025b) The Photochemistry of Amino Acids Produced on the Polar Cryovolcanic Regions of Titan. ACS Earth Space Chem 9:715–728. https://doi.org/10.1021/ACSEARTHSPACECHEM.4C00376

Gonçalves D, Ribeiro C, Tomé JPC, et al (2026) Towards fiber optic chemical sensors for 1,3-butadiene detection in Titan's hydrocarbon lakes. Anal Chim Acta 1384:344982. https://doi.org/10.1016/J.ACA.2025.344982

Goossens S, van Noort B, Mateo A, et al (2024) A low-density ocean inside Titan inferred from Cassini data. Nat Astron 8:846–855. https://doi.org/10.1038/S41550-024-02253-4

Grubisic A, Trainer MG, Li X, et al (2021) Laser Desorption Mass Spectrometry at Saturn's moon Titan. Int J Mass Spectrom 470:116707. https://doi.org/10.1016/J.IJMS.2021.116707

Hadraoui K, Cottin H, Ivanovski SL, et al (2019) Distributed glycine in comet 67P/Churyumov-Gerasimenko. Astron Astrophys 630:A32. https://doi.org/10.1051/0004-6361/201935018

Hand KP, Sotin C, Hayes A, Coustenis A (2020) On the Habitability and Future Exploration of Ocean Worlds. Space Sci Rev 216:1–24. https://doi.org/10.1007/s11214-020-00713-7

Hayes A, Aharonson O, Callahan P, et al (2008) Hydrocarbon lakes on Titan: Distribution and interaction with a porous regolith. Geophys Res Lett 35:L09204. https://doi.org/10.1029/2008GL033409

Hayes AG, Birch SPD, Dietrich WE, et al (2017) Topographic Constraints on the Evolution and Connectivity of Titan's Lacustrine Basins. Geophys Res Lett 44:11,745-11,753. https://doi.org/10.1002/2017GL075468

Hayes AG, Lorenz RD, Lunine JI (2018) A post-Cassini view of Titan's methane-based hydrologic cycle. Nat Geosci 11:306–313. https://doi.org/10.1038/s41561-018-0103-y

He C, Hörst SM, Radke M, Yant M (2022) Optical Constants of a Titan Haze Analog from 0.4 to 3.5 μm Determined Using Vacuum Spectroscopy. Planet Sci J 3:25. https://doi.org/10.3847/PSJ/AC4793

He C, Smith MA (2014) Identification of nitrogenous organic species in Titan aerosols analogs: Implication for prebiotic chemistry on Titan and early Earth. Icarus 238:86–92. https://doi.org/10.1016/J.ICARUS.2014.05.012

Hedgepeth JE, Buffo JJ, Chivers CJ, et al (2022) Modeling the Distribution of Organic Carbon and Nitrogen in Impact Crater Melt on Titan. Planet Sci J 3:51. https://doi.org/10.3847/PSJ/AC4D9C

Hedgepeth JE, Neish CD, Turtle EP, et al (2020) Titan's impact crater population after Cassini. Icarus 344:113664. https://doi.org/10.1016/J.ICARUS.2020.113664

Hemingway D, Nimmo F, Zebker H, Iess L (2013) A rigid and weathered ice shell on Titan. Nature 500:550–552. https://doi.org/10.1038/nature12400

Hendrix AR, Hurford TA, Barge LM, et al (2019) The NASA Roadmap to Ocean Worlds. Astrobiology 19:1–27. https://doi.org/10.1089/AST.2018.1955

Hersant F, Gautier D, Tobie G, Lunine JI (2008) Interpretation of the carbon abundance in Saturn measured by Cassini. Planet Space Sci 56:1103–1111. https://doi.org/10.1016/J.PSS.2008.02.007

Hörst SM (2017) Titan's atmosphere and climate. J Geophys Res Planets 122:432–482. https://doi.org/10.1002/2016JE005240

Hörst SM, He C, Ugelow MS, et al (2018) Exploring the Atmosphere of Neoproterozoic Earth: The Effect of O2 on Haze Formation and Composition. Astrophys J 858:119. https://doi.org/10.3847/1538-4357/AABD7D

Hörst SM, Vuitton V, Yelle R V. (2008) Origin of oxygen species in Titan's atmosphere. J Geophys Res 113:E10006. https://doi.org/10.1029/2008JE003135

Hörst SM, Yelle RV, Buch A, et al (2012) Formation of Amino Acids and Nucleotide Bases in a Titan Atmosphere Simulation Experiment. Astrobiology 12:809–817. https://doi.org/10.1089/AST.2011.0623

Iess L, Jacobson RA, Ducci M, et al (2012) The tides of Titan. Science (1979) 337:457–459. https://doi.org/10.1126/science.1219631

Iess L, Rappaport NJ, Jacobson RA, et al (2010) Gravity Field, Shape, and Moment of Inertia of Titan. Science (1979) 327:1367–1369. https://doi.org/10.1126/SCIENCE.1182583

Izquierdo-Ruiz F, Cable ML, Hodyss R, et al (2025) Hydrogen cyanide and hydrocarbons mix on Titan. Proc Natl Acad Sci U S A 122:e2507522122. https://doi.org/10.1073/PNAS.2507522122

Jansen KT, Reed NW, Browne EC, Tolbert MA (2025) Measurement of Photochemical Haze Refractive Indices and Hygroscopicity: Influence of CO2 in CH4/H2S/N2 Mixtures. Astrobiology 25:395–403. https://doi.org/10.1089/AST.2024.0142

Janssen MA, Le Gall A, Lopes RM, et al (2016) Titan's surface at 2.18-cm wavelength imaged by the Cassini RADAR radiometer: Results and interpretations through the first ten years of observation. Icarus 270:443–459. https://doi.org/10.1016/j.icarus.2015.09.027

Jennings DE, Tokano T, Cottini V, et al (2019) Titan Surface Temperatures during the Cassini Mission. Astrophys J Lett 877:L8. https://doi.org/10.3847/2041-8213/AB1F91

Johnson P V., Hodyss R, Malaska MJ (2025) Infrared Spectra of Organic and Salt-bearing Ices under Titan Conditions: Constraints on Detection of Ocean Materials. Planet Sci J 6:274. https://doi.org/10.3847/PSJ/AE1733

Journaux B, Daniel I, Petitgirard S, et al (2017) Salt partitioning between water and high-pressure ices. Implication for the dynamics and habitability of icy moons and water-rich planetary bodies. Earth Planet Sci Lett 463:36–47. https://doi.org/10.1016/J.EPSL.2017.01.017

Journaux B, Kalousová K, Sotin C, et al (2020) Large Ocean Worlds with High-Pressure Ices. Space Sci Rev 216:7-. https://doi.org/10.1007/S11214-019-0633-7

Kalousová K, Sotin C (2018) Melting in High-Pressure Ice Layers of Large Ocean Worlds—Implications for Volatiles Transport. Geophys Res Lett 45:8096–8103. https://doi.org/10.1029/2018GL078889

Kalousová K, Sotin C (2020) The Insulating Effect of Methane Clathrate Crust on Titan's Thermal Evolution. Geophys Res Lett 47:e2020GL087481. https://doi.org/10.1029/2020GL087481

Kalousová K, Wakita S, Sotin C, et al (2024) Evolution of Impact Melt Pools on Titan. J Geophys Res Planets 129:e2023JE008107. https://doi.org/10.1029/2023JE008107

Kargel JS (1991) Brine volcanism and the interior structures of asteroids and icy satellites. Icarus 94:368–390. https://doi.org/10.1016/0019-1035(91)90235-L

Kargel JS (1992) Ammonia-water volcanism on icy satellites: Phase relations at 1 atmosphere. Icarus 100:556–574. https://doi.org/10.1016/0019-1035(92)90118-Q

Khare BN, Sagan C, Ogino H, et al (1986) Amino acids derived from Titan Tholins. Icarus 68:176–184. https://doi.org/10.1016/0019-1035(86)90080-1

Khawaja N, Postberg F, O'Sullivan TR, et al (2025) Detection of organic compounds in freshly ejected ice grains from Enceladus's ocean. Nat Astron 9:1662–1671. https://doi.org/10.1038/s41550-025-02655-y

Lapôtre MGA, Malaska MJ, Cable ML (2022) The Role of Seasonal Sediment Transport and Sintering in Shaping Titan's Landscapes: A Hypothesis. Geophys Res Lett 49:e2021GL097605. https://doi.org/10.1029/2021GL097605

Lawrence DJ, Adams D, Ames WF, et al (2022) The Dragonfly Gamma-Ray and Neutron Spectrometer (DraGNS): Providing In-situ Measurements of Titan's Elemental Composition. In: 53rd Lunar and Planetary Science Conference. p 1939

Lebonnois S, Bakes ELO, McKay CP (2002) Transition from Gaseous Compounds to Aerosols in Titan's Atmosphere. Icarus 159:505–517. https://doi.org/10.1006/ICAR.2002.6943

Leitner MA, Lunine JI (2019) Modeling early Titan's ocean composition. Icarus 333:61–70. https://doi.org/10.1016/j.icarus.2019.05.008

Li J, Yu X, Sciamma-O'Brien E, et al (2022) A Cross-laboratory Comparison Study of Titan Haze Analogs: Surface Energy. Planet Sci J 3:2. https://doi.org/10.3847/PSJ/AC3D27

Lombardo NA, Lora JM (2023) The Heat and Momentum Budgets of Titan's Middle Atmosphere. J Geophys Res Planets 128:e2023JE008061. https://doi.org/10.1029/2023JE008061

Lopes RMC, Elachi C, Müeller-Wodarg ICF, Solomonidou A (2025) Titan After Cassini-Huygens. Elsevier

Lopes RMC, Kirk RL, Mitchell KL, et al (2013) Cryovolcanism on Titan: New results from Cassini RADAR and VIMS. Journal of Geophysical Research E: Planets 118:416–435. https://doi.org/10.1002/jgre.20062

Lopes RMC, Malaska MJ, Schoenfeld AM, et al (2020) A global geomorphologic map of Saturn's moon Titan. Nat Astron 4:228–233. https://doi.org/10.1038/s41550-019-0917-6

Lopes RMC, Mitchell KL, Stofan ER, et al (2007) Cryovolcanic features on Titan's surface as revealed by the Cassini Titan Radar Mapper. Icarus 186:395–412. https://doi.org/10.1016/j.icarus.2006.09.006

Lorenz RD, Le Gall A (2020) Schumann resonance on Titan: A critical Re-assessment. Icarus 351:113942. https://doi.org/10.1016/J.ICARUS.2020.113942

Lorenz RD, MacKenzie SM, Neish CD, et al (2021) Selection and Characteristics of the Dragonfly Landing Site near Selk Crater, Titan. Planet Sci J 2:24. https://doi.org/10.3847/PSJ/ABD08F

Lorenz RD, Oleson SR, Colozza AJ, et al (2018a) Exploring Titan's cryogenic hydrocarbon seas with boat-deployed expendable dropsondes. Advances in Space Research 62:912–920. https://doi.org/10.1016/J.ASR.2018.05.030

Lorenz RD, Turtle EP, Barnes JW, et al (2018b) Dragonfly: A rotorcraft lander concept for scientific exploration at titan. Johns Hopkins APL Technical Digest (Applied Physics Laboratory) 34:374–387

Lunine JI (2017) Ocean worlds exploration. Acta Astronaut 131:123–130. https://doi.org/10.1016/j.actaastro.2016.11.017

Lunine JI (2009) Saturn's Titan: A Strict Test for Life's Cosmic Ubiquity. Proc Am Philos Soc 153:403–418

Lunine JI, Cable ML, Horst SM, Rahm M (2020) The Astrobiology of Titan. In: Meadows VS, Arney GN, Schmidt BE, Des Marais DJ (eds) Planetary Astrobiology. University of Arizona Press, pp 247–266

Lunine JI, Lorenz RD (2009) Rivers, Lakes, Dunes, and Rain: Crustal Processes in Titan's Methane Cycle. Annu Rev Earth Planet Sci 37:299–320. https://doi.org/10.1146/annurev.earth.031208.100142

Lv KP, Norman L, Li YL (2017) Oxygen-Free Biochemistry: The Putative CHN Foundation for Exotic Life in a Hydrocarbon World? Astrobiology 17:1173–1181. https://doi.org/10.1089/AST.2016.1574

MacKenzie SM, Birch SPD, Hörst S, et al (2021) Titan: Earth-like on the Outside, Ocean World on the Inside. Planet Sci J 2:112. https://doi.org/10.3847/PSJ/ABF7C9

MacKenzie SM, Pontefract A, Daly RT, et al (2024) Impacts on Ocean Worlds Are Sufficiently Frequent and Energetic to Be of Astrobiological Importance. Planet Sci J 5:176. https://doi.org/10.3847/PSJ/AD656B

Madan I, Pearce BKD (2026) Prebiotic Chemistry Insights for Dragonfly. II. Thermodynamic Favorability of Nucleobases, Ribose, and Fatty Acids in Selk Crater on Titan. Planet Sci J 7:107. https://doi.org/10.3847/PSJ/AE5F91

Madan I, Pearce BKD (2025) Prebiotic Chemistry Insights for Dragonfly: Thermodynamics of Amino Acid Synthesis in Selk Crater on Titan. Planet Sci J 6:284. https://doi.org/10.3847/PSJ/AE1C18

Malaska MJ, Hodyss R (2014) Dissolution of benzene, naphthalene, and biphenyl in a simulated Titan lake. Icarus 242:74–81. https://doi.org/10.1016/J.ICARUS.2014.07.022

Malaska MJ, Radebaugh J, Lopes RMC, et al (2020) Labyrinth terrain on Titan. Icarus 344:113764. https://doi.org/10.1016/J.ICARUS.2020.113764

Malaska MJ, Schoenfeld A, Wynne JJ, et al (2022) Potential Caves: Inventory of Subsurface Access Points on the Surface of Titan. J Geophys Res Planets 127:e2022JE007512. https://doi.org/10.1029/2022JE007512

Mandt K, Luspay-Kuti A, Lustig-Yaeger J, et al (2022) TRAPPIST-1h as an Exo-Titan. I. The Role of Assumptions about Atmospheric Parameters in Understanding an Exoplanet Atmosphere. Astrophys J 930:73. https://doi.org/10.3847/1538-4357/AC59BB

Martínez-Rodríguez H, Caballero JA, Cifuentes C, et al (2019) Exomoons in the Habitable Zones of M Dwarfs. Astrophys J 887:261. https://doi.org/10.3847/1538-4357/AB5640

Martins Z, Bunce E, Grasset O, et al (2024) Report of the Expert Committee for the Large-class mission in ESA's Voyage 2050 plan covering the science theme "Moons of the Giant Planets"

Martins Z, Pasek MA (2024) Delivery of Organic Matter to the Early Earth. Elements 20:19–23. https://doi.org/10.2138/GSELEMENTS.20.1.19

Martins Z, Price MC, Goldman N, et al (2013) Shock synthesis of amino acids from impacting cometary and icy planet surface analogues. Nat Geosci 6:1045–1049. https://doi.org/10.1038/ngeo1930

Mayer C, Nixon CA (2025) A proposed mechanism for the formation of protocell-like structures on Titan. Int J Astrobiol 24:e7. https://doi.org/10.1017/S1473550425100037

McKay C (2016) Titan as the Abode of Life. 6:8. https://doi.org/10.3390/life6010008

McKay CP (2014) Requirements and limits for life in the context of exoplanets. Proc Natl Acad Sci U S A 111:12628–12633. https://doi.org/10.1073/pnas.1304212111

McKay CP, Pollack JB, Courtin R (1991) The greenhouse and antigreenhouse effects on Titan. Science (1979) 253:1118–1121. https://doi.org/10.1126/science.11538492

McKay CP, Smith HD (2005) Possibilities for methanogenic life in liquid methane on the surface of Titan. Icarus 178:274–276. https://doi.org/10.1016/J.ICARUS.2005.05.018

Miller KE, Glein CR, Waite JH (2019) Contributions from Accreted Organics to Titan's Atmosphere: New Insights from Cometary and Chondritic Data. Astrophys J 871:59. https://doi.org/10.3847/1538-4357/AAF561

Miller KE, Melwani Daswani M, Sotin C, et al (2026) Titan's refractory core evolution: Implications for organics in its subsurface ocean. Icarus 449:116961. https://doi.org/10.1016/J.ICARUS.2026.116961

Mitchell JL, Lora JM (2016) The Climate of Titan. Annu Rev Earth Planet Sci 44:353–380. https://doi.org/10.1146/ANNUREV-EARTH-060115-012428

Mitri G, Barnes J, Coustenis A, et al (2021) Exploration of Enceladus and Titan: investigating ocean worlds' evolution and habitability in the Saturn system. Exp Astron (Dordr) 54:877–910. https://doi.org/10.1007/S10686-021-09772-2

Mitri G, Coustenis A, Fanchini G, et al (2014a) The exploration of Titan with an orbiter and a lake probe. Planet Space Sci 104:78–92. https://doi.org/10.1016/J.PSS.2014.07.009

Mitri G, Lunine JI, Mastrogiuseppe M, Poggiali V (2019) Possible explosion crater origin of small lake basins with raised rims on Titan. Nat Geosci 12:791–796. https://doi.org/10.1038/s41561-019-0429-0

Mitri G, Meriggiola R, Hayes A, et al (2014b) Shape, topography, gravity anomalies and tidal deformation of Titan. Icarus 236:169–177. https://doi.org/10.1016/j.icarus.2014.03.018

Mitri G, Showman AP (2008) Thermal convection in ice-I shells of Titan and Enceladus. Icarus 193:387–396. https://doi.org/10.1016/J.ICARUS.2007.07.016

Mitri G, Showman AP, Lunine JI, Lopes RMC (2008) Resurfacing of Titan by ammonia-water cryomagma. Icarus 196:216–224. https://doi.org/10.1016/j.icarus.2008.02.024

Moore JM, Pappalardo RT (2011) Titan: An exogenic world? Icarus 212:790–806. https://doi.org/10.1016/J.ICARUS.2011.01.019

Morrison AA, Whittington AG, Zhong F, et al (2022) The viscosity of aqueous solutions as analogs to cryovolcanic liquids. Geochim Cosmochim Acta 339:97–114. https://doi.org/10.1016/J.GCA.2022.10.019

Mortensen D, Cranston BC, Pergosky DD, et al (2024a) The Dragonfly Surface Thermal Properties Sensor (STHERM): Design and Testing. In: IEEE Aerospace Conference Proceedings. IEEE

Mortensen D, Shackford JA, Lowery KR, et al (2024b) Development and Testing of the Dragonfly Geophysics and Meteorology (DraGMet) WIND Sensor. In: IEEE Aerospace Conference Proceedings. IEEE

Moulay V, Freissinet C, Rizk-Bigourd M, et al (2023) Selection and Analytical Performances of the Dragonfly Mass Spectrometer Gas Chromatographic Columns to Support the Search for Organic Molecules of Astrobiological Interest on Titan. Astrobiology 23:213–229. https://doi.org/10.1089/AST.2022.0038

Munoz-Iglesias V, Prieto-Ballesteros O (2022) Low-Temperature High-Pressure Chemistry of Ammonia and Methanol Aqueous Solutions in the Presence of Different Carbon Sources: Application to Icy Bodies. ACS Earth Space Chem 6:1482–1494. https://doi.org/10.1021/ACSEARTHSPACECHEM.1C00407

Nasralla M, Laurent H, Alderman OLG, Dougan L (2025) Solution structure of Titan-relevant aqueous ammonia by neutron diffraction. Commun Chem 8:227-. https://doi.org/10.1038/s42004-025-01599-8

Neish CD, Kirk RL, Lorenz RD, et al (2013) Crater topography on Titan: Implications for landscape evolution. Icarus 223:82–90. https://doi.org/10.1016/J.ICARUS.2012.11.030

Neish CD, Lorenz RD (2012) Titan's global crater population: A new assessment. Planet Space Sci 60:26–33. https://doi.org/10.1016/J.PSS.2011.02.016

Neish CD, Lorenz RD, O'Brien DP (2006) The potential for prebiotic chemistry in the possible cryovolcanic dome Ganesa Macula on Titan. Int J Astrobiol 5:57–65. https://doi.org/10.1017/S1473550406002898

Neish CD, Lorenz RD, Turtle EP, et al (2018) Strategies for Detecting Biological Molecules on Titan. Astrobiology 18:571–585. https://doi.org/10.1089/AST.2017.1758

Neish CD, Malaska MJ, Sotin C, et al (2024) Organic Input to Titan's Subsurface Ocean Through Impact Cratering. Astrobiology 24:177–189. https://doi.org/10.1089/AST.2023.0055

Neish CD, Molaro JL, Lora JM, et al (2016) Fluvial erosion as a mechanism for crater modification on Titan. Icarus 270:114–129. https://doi.org/10.1016/J.ICARUS.2015.07.022

Neish CD, Somogyi Á, Imanaka H, et al (2008) Rate measurements of the hydrolysis of complex organic macromolecules in cold aqueous solutions: Implications for prebiotic chemistry on the early earth and titan. Astrobiology 8:273–287. https://doi.org/10.1089/AST.2007.0193

Neish CD, Somogyi Á, Lunine JI, Smith MA (2009) Low temperature hydrolysis of laboratory tholins in ammonia-water solutions: Implications for prebiotic chemistry on Titan. Icarus 201:412–421. https://doi.org/10.1016/j.icarus.2009.01.003

Neish CD, Somogyi Á, Smith MA (2010) Titan's primordial soup: Formation of amino acids via low-temperature hydrolysis of tholins. Astrobiology 10:337–347. https://doi.org/10.1089/AST.2009.0402

Néri A, Guyot F, Reynard B, Sotin C (2020) A carbonaceous chondrite and cometary origin for icy moons of Jupiter and Saturn. Earth Planet Sci Lett 530:115920. https://doi.org/10.1016/j.epsl.2019.115920

Neveu M, Desch SJ, Shock EL, Glein CR (2015) Prerequisites for explosive cryovolcanism on dwarf planet-class Kuiper belt objects. Icarus 246:48–64. https://doi.org/10.1016/J.ICARUS.2014.03.043

Niemann HB, Atreya SK, Bauer SJ, et al (2005) The abundances of constituents of Titan's atmosphere from the GCMS instrument on the Huygens probe. Nature 438:779–784. https://doi.org/10.1038/nature04122

Niemann HB, Atreya SK, Demick JE, et al (2010) Composition of Titan's lower atmosphere and simple surface volatiles as measured by the Cassini-Huygens probe gas chromatograph mass spectrometer experiment. Journal of Geophysical Research E: Planets 115:12006. https://doi.org/10.1029/2010JE003659

Niraula P, Wit J de, Hargreaves RJ, et al (2025) The Detection-versus-retrieval Challenge: Titan as an Exoplanet. Astrophys J Lett 995:L66. https://doi.org/10.3847/2041-8213/AE27C2

Nixon CA (2024) The Composition and Chemistry of Titan's Atmosphere. ACS Earth Space Chem 8:406–456. https://doi.org/10.1021/ACSEARTHSPACECHEM.2C00041

Nixon CA, Ansty TM, Lombardo NA, et al (2019) Cassini Composite Infrared Spectrometer (CIRS) Observations of Titan 2004–2017. Astrophys J Suppl Ser 244:14. https://doi.org/10.3847/1538-4365/AB3799

Nixon CA, Birch S, Chatain A, et al (2026) Terrestrial Analogs to Titan for Geophysical Research. Reviews of Geophysics 64:e2025RG000909. https://doi.org/10.1029/2025RG000909

Nixon CA, Carrasco N, Sotin C (2025) Open questions and future directions in Titan science. In: Lopes RMC, Elachi C, Müeller-Wodarg ICF, Solomonidou A (eds) Titan after Cassini-Huygens. Elsevier, pp 473–515

O'Brien DP, Lorenz RD, Lunine JI (2005) Numerical calculations of the longevity of impact oases on Titan. Icarus 173:243–253. https://doi.org/10.1016/J.ICARUS.2004.08.001

Pasek MA, Mousis O, Lunine JI (2011) Phosphorus chemistry on Titan. Icarus 212:751–761. https://doi.org/10.1016/J.ICARUS.2011.01.026

Pearce BKD, Hörst SM, Cline CJ, et al (2024a) Toward Prebiotic Chemistry on Titan: Impact Experiments on Organic Haze Particles. Planet Sci J 5:68. https://doi.org/10.3847/PSJ/AD283E

Pearce BKD, Hörst SM, Sebree JA, He C (2024b) Organic Hazes as a Source of Life's Building Blocks to Warm Little Ponds on the Hadean Earth. Planet Sci J 5:23. https://doi.org/10.3847/PSJ/AD17BD

Petricca F, Vance SD, Parisi M, et al (2025) Titan's strong tidal dissipation precludes a subsurface ocean. Nature 648:556–561. https://doi.org/10.1038/s41586-025-09818-x

Poch O, Coll P, Buch A, et al (2012) Production yields of organics of astrobiological interest from H2O–NH3 hydrolysis of Titan's tholins. Planet Space Sci 61:114–123. https://doi.org/10.1016/J.PSS.2011.04.009

Postberg F, Sekine Y, Klenner F, et al (2023) Detection of phosphates originating from Enceladus's ocean. Nature 618:489–493. https://doi.org/10.1038/s41586-023-05987-9

Purkamo L, Bomberg M, Nyyssönen M, et al (2014) Heterotrophic Communities Supplied by Ancient Organic Carbon Predominate in Deep Fennoscandian Bedrock Fluids. Microb Ecol 69:319–332. https://doi.org/10.1007/S00248-014-0490-6

Ramírez SI, Coll P, Buch A, et al (2010) The fate of aerosols on the surface of Titan. Faraday Discuss 147:419–427. https://doi.org/10.1039/C003925J

Ranjan S, Wogan NF, Glidden A, et al (2025) The Photochemical Plausibility of Warm Exo-Titans Orbiting M Dwarf Stars. Astrophys J Lett 993:L39. https://doi.org/10.3847/2041-8213/AE1026

Read PL, Barstow J, Charnay B, et al (2016) Global energy budgets and "Trenberth diagrams" for the climates of terrestrial and gas giant planets. Quarterly Journal of the Royal Meteorological Society 142:703–720. https://doi.org/10.1002/qj.2704

Reynard B, Sotin C (2023) Carbon-rich icy moons and dwarf planets. Earth Planet Sci Lett 612:118172. https://doi.org/10.1016/J.EPSL.2023.118172

Rianço-Silva R, Machado P, Martins Z, et al (2024) A study of very high resolution visible spectra of Titan: Line characterisation in visible CH4 bands and the search for C3. Planet Space Sci 240:105836. https://doi.org/10.1016/J.PSS.2023.105836

Rianço-Silva R, Machado P, Rannou P, et al (2026) Detection of C3 in Titan with VLT-ESPRESSO. Mon Not R Astron Soc 548:. https://doi.org/10.1093/MNRAS/STAG447

Rodriguez S, Vinatier S, Cordier D, et al (2022) Science goals and new mission concepts for future exploration of Titan's atmosphere, geology and habitability: titan POlar scout/orbitEr and in situ lake lander and DrONe explorer (POSEIDON). Exp Astron (Dordr) 54:911–973. https://doi.org/10.1007/S10686-021-09815-8

Rossignoli NL, Di Sisto RP, Parisi MG (2022) Crater production on Titan and surface chronology. Astron Astrophys 660:A127. https://doi.org/10.1051/0004-6361/202141802

Sagan C, Khare BN (1979) Tholins: organic chemistry of interstellar grains and gas. Nature 277:102–107. https://doi.org/10.1038/277102a0

Sandström H, Rahm M (2020) Can polarity-inverted membranes self-assemble on Titan? Sci Adv 6:. https://doi.org/10.1126/SCIADV.AAX0272

Schoenfeld AM, Solomonidou A, Malaska MJ, et al (2023) Geomorphological Map of the Soi Crater Region on Titan. J Geophys Res Planets 128:e2022JE007499. https://doi.org/10.1029/2022JE007499

Schurmeier LR, Brouwer GE, Kay JP, et al (2024) Rapid Impact Crater Relaxation Caused by an Insulating Methane Clathrate Crust on Titan. Planet Sci J 5:211. https://doi.org/10.3847/PSJ/AD7018

Schurmeier LR, Dombard AJ (2018) Crater relaxation on Titan aided by low thermal conductivity sand infill. Icarus 305:314–323. https://doi.org/10.1016/J.ICARUS.2017.10.034

Schurmeier LR, Dombard AJ, Malaska MJ, et al (2023) An intrusive cryomagmatic origin for northern radial labyrinth terrains on Titan and implications for the presence of crustal clathrates. Icarus 404:115664. https://doi.org/10.1016/J.ICARUS.2023.115664

Sciamma-O’Brien E, Roush TL, Rannou P, et al (2023) First Optical Constants of Laboratory-generated Organic Refractory Materials (Tholins) Produced in the NASA Ames COSmIC Facility from the Visible to the Near Infrared (0.4–1.6 μm): Application to Titan’s Aerosols. Planet Sci J 4:121. https://doi.org/10.3847/PSJ/ACD83F

Sebree JA, Roach MC, Shipley ER, et al (2018) Detection of Prebiotic Molecules in Plasma and Photochemical Aerosol Analogs Using GC/MS/MS Techniques. Astrophys J 865:133. https://doi.org/10.3847/1538-4357/AADBA1

Seltzer C, Martel SJ, Perron JT (2025) Topographic Stress as a Mechanical Weathering Mechanism on Titan. J Geophys Res Planets 130:e2024JE008873. https://doi.org/10.1029/2024JE008873

Seltzer C, O’Ghaffari H, Peč M (2024) Mechanical Properties of LL6 Chondrites Under Pressures Relevant to Rocky Interiors of Icy Moons. J Geophys Res Planets 129:e2024JE008296. https://doi.org/10.1029/2024JE008296

Shah J, Neish CD, Trozzo S (2025) An analogue study of impact craters on Titan: Implications for Titan’s surface age. Icarus 433:116536. https://doi.org/10.1016/J.ICARUS.2025.116536

Sithajan S, Kaewbiang L, Jones HRA, et al (2025) Identification of Likely Methane Absorption Features in the Optical Spectra of Titan. Astron J 170:315. https://doi.org/10.3847/1538-3881/AE0E1F

Soderblom JM, Brown RH, Soderblom LA, et al (2010) Geology of the Selk crater region on Titan from Cassini VIMS observations. Icarus 208:905–912. https://doi.org/10.1016/J.ICARUS.2010.03.001

Sohl F, Solomonidou A, Wagner FW, et al (2014) Structural and tidal models of Titan and inferences on cryovolcanism. J Geophys Res Planets 119:1013–1036. https://doi.org/10.1002/2013JE004512

Solomonidou A, Coustenis A, Hirtzig M, et al (2016) Temporal variations of Titan’s surface with Cassini/VIMS. Icarus 270:85–99. https://doi.org/10.1016/J.ICARUS.2015.05.003

Solomonidou A, Coustenis A, Lopes RMC, et al (2018) The Spectral Nature of Titan’s Major Geomorphological Units: Constraints on Surface Composition. J Geophys Res Planets 123:489–507. https://doi.org/10.1002/2017JE005477

Solomonidou A, Neish C, Coustenis A, et al (2020) The chemical composition of impact craters on Titan - I. Implications for exogenic processing. Astron Astrophys 641:A16. https://doi.org/10.1051/0004-6361/202037866

Sotin C, Kalousovaacute K, Tobie G (2021) Titan’s interior structure and dynamics after the Cassini-Huygens mission. Annu Rev Earth Planet Sci 49:579–607. https://doi.org/10.1146/ANNUREV-EARTH-072920-052847

Stevenson J, Lunine J, Clancy P (2015) Membrane alternatives in worlds without oxygen: Creation of an azotosome. Sci Adv 1:e1400067. https://doi.org/10.1126/sciadv.1400067

Stofan E, Lorenz R, Lunine J, et al (2013) TiME-the titan mare explorer. In: IEEE Aerospace Conference Proceedings. IEEE

Strobel DF (2010) Molecular hydrogen in Titan's atmosphere: Implications of the measured tropospheric and thermospheric mole fractions. Icarus 208:878–886. https://doi.org/10.1016/J.ICARUS.2010.03.003

Strobel DF (2022) Molecular hydrogen in the upper atmospheres of Saturn and Titan. Icarus 376:114876. https://doi.org/10.1016/J.ICARUS.2021.114876

Tagliabue A, Schneider S, Pavone M, Agha-Mohammadi AA (2020) Shapeshifter: A Multi-Agent, Multi-Modal Robotic Platform for Exploration of Titan. In: IEEE Aerospace Conference Proceedings. IEEE

Taniuchi T, Takano Y, Kobayashi K (2013) Amino Acid Precursors from a Simulated Lower Atmosphere of Titan: Experiments of Cosmic Ray Energy Source with 13C- and 18O-Stable Isotope Probing Mass Spectrometry. Analytical Sciences 29:777–785. https://doi.org/10.2116/ANALSCI.29.777/METRICS

Thakur AC, Remsing RC (2024) Molecular Structure and Rotational Dynamics in the Acetonitrile:Acetylene (1:2) Plastic Co-crystal at Titan Conditions. ACS Earth Space Chem 8:1661–1667. https://doi.org/10.1021/ACSEARTHSPACECHEM.4C00157

Thakur AC, Remsing RC (2023) Molecular Structure, Dynamics, and Vibrational Spectroscopy of the Acetylene:Ammonia (1:1) Plastic Co-Crystal at Titan Conditions. ACS Earth Space Chem 7:479–489. https://doi.org/10.1021/ACSEARTHSPACECHEM.2C00327

Tinetti G, Drossart P, Eccleston P, et al (2018) A chemical survey of exoplanets with ARIEL. Exp Astron (Dordr) 46:135–209. https://doi.org/10.1007/S10686-018-9598-X

Tobie G, Gautier D, Hersant F (2012) Titan's bulk composition constrained by Cassini–Huygens: implication for internal outgassing. Astrophys J 752:125. https://doi.org/10.1088/0004-637X/752/2/125

Tobie G, Grasset O, Lunine JI, et al (2005) Titan's internal structure inferred from a coupled thermal-orbital model. Icarus 175:496–502. https://doi.org/10.1016/J.ICARUS.2004.12.007

Tobie G, Lunine JI, Sotin C (2006) Episodic outgassing as the origin of atmospheric methane on Titan. Nature 440:61–64. https://doi.org/10.1038/nature04497

Tomasko MG, Archinal B, Becker T, et al (2005) Rain, winds and haze during the Huygens probe's descent to Titan's surface. Nature 438:765–778. https://doi.org/10.1038/nature04126

Trainer MG, Pavlov AA, DeWitt HL, et al (2006) Organic haze on Titan and the early Earth. Proc Natl Acad Sci U S A 103:18035–18042. https://doi.org/10.1073/PNAS.0608561103

Ugelow MS, Wieman ST, Schwarz MCR, et al (2024) Laboratory Studies on the Influence of Hydrogen on Titan-like Photochemistry. ACS Earth Space Chem 8:2362–2371. https://doi.org/10.1021/ACSEARTHSPACECHEM.4C00102

Vance SD, Panning MP, Stähler S, et al (2018) Geophysical Investigations of Habitability in Ice-Covered Ocean Worlds. J Geophys Res Planets 123:180–205. https://doi.org/10.1002/2017JE005341

Vu TH, Choukroun M, Sotin C, et al (2020) Rapid Formation of Clathrate Hydrate From Liquid Ethane and Water Ice on Titan. Geophys Res Lett 47:e2019GL086265. https://doi.org/10.1029/2019GL086265

Vu TH, Hodyss R (2026) Experimental insights into the azotosome hypothesis in Titan's lake fluids. Sci Adv 12:eaed1426. https://doi.org/10.1126/SCIADV.AED1426

Waite JH, Lewis WS, Magee BA, et al (2009) Liquid water on Enceladus from observations of ammonia and 40Ar in the plume. Nature 460:487–490. https://doi.org/10.1038/nature08153

Wakita S, Johnson BC, Soderblom JM, et al (2023) Modeling the Formation of Selk Impact Crater on Titan: Implications for Dragonfly. Planet Sci J 4:51. https://doi.org/10.3847/PSJ/ACBE40

Wakita S, Johnson BC, Soderblom JM, et al (2022) Methane-saturated Layers Limit the Observability of Impact Craters on Titan. Planet Sci J 3:50. https://doi.org/10.3847/PSJ/AC4E91

Wakita S, Johnson BC, Soderblom JM, Neish CD (2026) A Youthful Titan Implied by Improved Impact Simulations. Geophys Res Lett 53:e2025GL116041. https://doi.org/10.1029/2025GL116041

Werynski A, Neish CD, Gall A Le, Janssen MA (2019) Compositional variations of Titan's impact craters indicates active surface erosion. Icarus 321:508–521. https://doi.org/10.1016/J.ICARUS.2018.12.007

Wilson EH, Atreya SK (2003) Chemical sources of haze formation in Titan's atmosphere. Planet Space Sci 51:1017–1033. https://doi.org/10.1016/J.PSS.2003.06.003

Wood CA, Lorenz R, Kirk R, et al (2010) Impact craters on Titan. Icarus 206:334–344. https://doi.org/10.1016/J.ICARUS.2009.08.021

Wood CA, Radebaugh J (2020) Morphologic Evidence for Volcanic Craters Near Titan's North Polar Region. J Geophys Res Planets 125:e2019JE006036. https://doi.org/10.1029/2019JE006036

Wray LB, Lorenz RD, Chapman KR, Huber CM (2024) Ears of the Dragonfly: Design of a Microphone System for Operation on Titan. In: IEEE Aerospace Conference Proceedings. IEEE

Yanez MD, LaRowe DE, Cable ML, Amend JP (2024) Energy yields for acetylenotrophy on Enceladus and Titan. Icarus 411:115969. https://doi.org/10.1016/J.ICARUS.2024.115969

Yang Z, Liu Y, He C, et al (2025) How CO Affects the Composition of Titan's Tholins Generated with Electron Cyclotron Resonance Plasma. Planet Sci J 6:47. https://doi.org/10.3847/PSJ/ADAEB5

Yu X, Yu Y, Garver J, et al (2024) The Fate of Simple Organics on Titan's Surface: A Theoretical Perspective. Geophys Res Lett 51:e2023GL106156. https://doi.org/10.1029/2023GL106156

Yu X, 余馨婷, Hörst SM, et al (2020) Surface Energy of the Titan Aerosol Analog "Tholin." Astrophys J 905:88. https://doi.org/10.3847/1538-4357/ABC55D

Zebker HA, Stiles B, Hensley S, et al (2009) Size and shape of saturn's moon titan. Science (1979) 324:921–923. https://doi.org/10.1126/SCIENCE.1168905

Zhong F, Mitchell KL, Hays CC, et al (2009) The rheology of cryovolcanic slurries: Motivation and phenomenology of methanol-water slurries with implications for Titan. Icarus 202:607–619. https://doi.org/10.1016/J.ICARUS.2009.03.015